\documentclass{aa}

\usepackage[utf8]{inputenc}
\usepackage[T1]{fontenc}
\usepackage{textcomp}
\usepackage{newunicodechar}
\newunicodechar{́}{\'}

\usepackage{graphicx}
\usepackage{svg}

\usepackage{txfonts}
\usepackage{xcolor}

\usepackage{placeins}

\usepackage{amsmath}
\usepackage{amssymb}

\usepackage{ulem}
\usepackage{makecell}

\usepackage{subcaption}

\usepackage{amsmath}

\usepackage{booktabs}
\usepackage{tabularx}

\usepackage[colorlinks=true,citecolor=blue,urlcolor=blue]{hyperref}
\defcitealias{Viscasillas2024}{Paper I}

\DeclareRobustCommand{\ion}[2]{%
\relax\ifmmode
\ifx\testbx\f@series
{\mathbf{#1\,\mathsc{#2}}}\else
{\mathrm{#1\,\mathsc{#2}}}\fi
\else\textup{#1\,{\mdseries\textsc{#2}}}%
\fi}

\begin{document} 

    \title{Detection of hot subdwarf binaries and sdB stars using machine learning methods and a large sample of Gaia XP~spectra}
    \titlerunning{short title}
    \authorrunning{name(s) of author(s)}

    \author{
            M. Ambrosch\inst{\ref{vilnius}} \and
            C. Viscasillas Vázquez\inst{\ref{vilnius}} \and 
            E. Solano\inst{\ref{cab}} \and 
            A. Ulla\inst{\ref{uvigo},\ref{ifcae}} \and
            X. Pérez-Couto\inst{\ref{udc},\ref{cigus1}} \and
            E. Pérez-Fernández\inst{\ref{uvigo}, \ref{beade}}  \and \\
            A. Medžiūnas \inst{\ref{vilnius2}} \and
            M. Manteiga\inst{\ref{cigus2}} \and
            C. Dafonte\inst{\ref{cigus1}} \and
            A. Drazdauskas\inst{\ref{vilnius}} \and
            L. Magrini\inst{\ref{oaa}} \and \\
            \v{S}. Mikolaitis\inst{\ref{vilnius}} \and
            V. \v{S}atas\inst{\ref{vilnius}}
           }

    \institute{
                Institute of Theoretical Physics and Astronomy, Vilnius University, Sauletekio av. 3, 10257 Vilnius, Lithuania. \label{vilnius}
                \and 
                Centro de Astrobiología (CSIC-INTA), Camino Bajo del Castillo s/n, E-28692 Villanueva de la Cañada, Madrid, Spain.  \label{cab}
                \and
                Applied Physics Department, Universidade de Vigo, Campus Lagoas-Marcosende, s/n, E-36310 Vigo, Spain. \label{uvigo}
                \and
                Centro de Investigación Mariña, Universidade de Vigo, CIM/GEOMA, 36310 Vigo, Spain \label{ifcae}
                \and
                Universidade da Coruña (UDC), Department of Computer Science and Information Technologies, Campus de Elviña s/n, 15071, A Coruña, Galiza, Spain. \label{udc}
                \and
                IES de Beade, Conseller ı́a de Educación e Ordenación Universitaria, Camino do Outeiro 10, E-36312 Vigo, Spain. \label{beade}
                \and
                Institute of Applied Mathematics, Vilnius University,  24 Naugarduko st., LT-03225 Vilnius, Lithuania. \label{vilnius2}
                \and
                CIGUS CITIC - Department of Computer Science and Information Technologies, University of A Coru\~na, s/n, E-15071 A Coru\~na, Spain. \label{cigus1}
                \and
                CIGUS CITIC - Department of Nautical Sciences and Marine Engineering, University of A Coru\~na, Paseo de Ronda 51, E-15011 A Coru\~na, Spain. \label{cigus2} \and 
                INAF - Osservatorio Astrofisico di Arcetri, Largo E. Fermi 5, 50125, Firenze, Italy. \label{oaa} 
            }

   \date{Received ; accepted }

   \abstract
    {Hot subdwarfs (hot sds) are compact, evolved stars near the Extreme Horizontal Branch (EHB), key to understanding stellar evolution and the UV excess in galaxies. We extend our previous analysis of Gaia XP spectra of hot subdwarf stars to a much larger sample, enabling a comprehensive study of their physical and binary properties.}
    {Our goal is to identify patterns in Gaia XP spectra, investigate binarity, and assess the influence of parameters such as temperature, helium abundance, and variability. We analyse $\sim$20{,}000 hot sd candidates selected from the literature, combining Gaia XP data with published parameters.}
    {We apply UMAP (Uniform Manifold Approximation and Projection) to the XP coefficients, which represent the Gaia XP spectra in a compact, feature-based form, to construct a similarity map. We then use self-organizing maps (SOMs) and convolutional neural networks (CNNs) to classify spectra as binaries or singles, and as cool/He-poor or hot/He-rich. The spectra are normalised using asymmetric least squares baseline fitting to emphasise individual spectral features.}
    {BP--RP colour dominates the similarity map, with additional influence from temperature, helium abundance, and variability. Most binaries, identified via the Virtual Observatory SED analyser (VOSA), cluster in two filaments linked to main-sequence companions. CNN classification suggests a strong correlation between variability and binarity, with binary fractions exceeding 60\% for active hot sds.}
    {Gaia XP spectra combined with dimensionality reduction and machine learning effectively reveal patterns in hot sd properties. Our findings indicate that binarity and environmental density strongly shape the evolutionary paths of hot subdwarfs. We identify possible contamination by main sequence and cataclysmic variable stars in our base sample.}
    {}


    \maketitle

\section{Introduction}
\label{section: Introduction}

Hot subdwarfs (hot sds) are compact, evolved stars located on or near the Extreme Horizontal Branch (EHB) of the Hertzsprung-Russell diagram (HRD). They are characterized by their high temperatures (20,000$-$50,000~K), low masses ($\sim$0.5~M$_{\odot}$), and hydrogen-deficient atmospheres. This unusual combination of parameters makes them crucial for understanding stellar evolution. Hot sds play a key role in explaining the ultraviolet (UV) excess observed in elliptical galaxies and bulges of spiral galaxies, and they provide insights into binary interactions such as mass transfer and common-envelope ejection. For a detailed discussion on the properties, formation channels, and observational significance of hot sds, refer to the introduction in \citet[][hereafter \citetalias{Viscasillas2024}]{Viscasillas2024} and references therein.

Binary interaction is thought to play an important role in the evolution of hot sds. They are found in binaries with a variety of companions, most commonly low-mass main-sequence stars and white dwarfs (WDs, e.g., \citealt{Pelisoli2021}), but also brown dwarfs (BDs, e.g., \citealt{Schaffenroth2021} and Be stars \citep[e.g.]{Mourard2015, Castanon2024}. 
Composite hot sds + main sequence (MS) systems with F$-$K type companions in wide orbits constitute a well-defined population \citep[e.g.]{Vos2018} and are the focus of this work. More rarely, hot sds systems have been proposed to host compact companions. While theoretical studies have explored systems with neutron-star companions \citep[e.g.]{Wu2018}, observational evidence remains scarce, with only two hot sds systems proposed as neutron-star candidates \citep{Mereghetti2021, Geier2023}; however, it is unclear whether the compact companions in these systems are neutron stars or massive white dwarfs. Claims of black-hole companions have also been reported \citep[e.g.]{Geier2010}, although their nature remains debated. In contrast, close hot sds+WD and hot sds+dM/BD binaries typically have orbital periods from tens of minutes to several days and generally do not show composite spectra. These systems reveal their binarity through photometric variability such as eclipses, reflection effects, or ellipsoidal modulation, but are too close to exhibit astrometric wobble or increased Gaia Renormalised Unit Weight Error (RUWE) values. The composite hot sds+MS binaries considered here are wide systems with orbital periods of a few hundred days, some of which show astrometric wobble and elevated RUWE values, but usually do not display binary-induced photometric variability; instead, their variability is dominated by sdB pulsations or weak rotational modulation of the cool companion \citep{Pelisoli2020}. Overall, both the relative frequencies of the different binary systems and their impact on hot sds evolution remain uncertain.

Astrophysics is one of the most data-intensive sciences, with missions like Gaia producing petabytes of information. Adopting advanced techniques for large-scale data analysis is essential for analysing these sheer amounts of data. Initiatives like the Virtual Observatory (VO) have been crucial in addressing challenges in massive data analysis. \citet{Oreiro2011} developed a VO-based procedure to identify previously uncatalogued hot sds while minimizing contamination from other stellar types. \citet{Perez-Fernandez2016} refined this method, successfully identifying new hot sds using hot sds spectra. \citet{Solano22} used the VOSA tool (Virtual Observatory SED Analyser, \citealt{Bayo2008}) to detect binary systems involving hot sds. There, binaries were identified by their flux excesses in the red part of the composite spectrum. Physical parameters of the hot sds and their companion stars could be estimated through spectral energy distribution (SED) fitting with VOSA. Building on these efforts, \citetalias{Viscasillas2024} introduced an advanced classification framework for hot sds binaries using AI techniques and Gaia~DR3 data. Their methodology combines supervised and unsupervised machine learning approaches, including Support Vector Machines (SVM), Self-Organizing Maps (SOM), Convolutional Neural Networks (CNN), the Uniform Manifold Approximation and Projection (UMAP) technique, and, for the first time for this purpose, the cosine similarity metric. Applying these techniques to 2815 hot subdwarfs with Gaia~DR3~BP/RP spectra, they achieved high classification accuracy. 

Other studies have also enhanced the classification of hot sds through the use of advanced techniques. \citet{Bu2019} and \cite{Tan2022} demonstrated the effectiveness of CNN and hybrid models for spectral analysis, achieving high precision in the identification of new hot sds candidates. The same or similar techniques have also been shown to be effective in classifying other classes of dwarf stars. Thus, most recently, \citet{Kao2024} used Gaia~DR3~BP/RP spectra and the UMAP technique to analyse $\sim$96,000 white dwarfs (WDs), identifying polluted WDs with multiple atmospheric metals. Similarly, \citet{PerezCouto2024} used SOMs to identify new polluted WD candidates among $\sim$66,000 WDs, showcasing their ability to detect heavy elements such as Ca, Mg, Na and K in cool WDs. Focusing on hot sds, \citet{Zhang2025} developed the Hot Subdwarf Detector (HsdDet), a multiscale object automated detection algorithm applied to SDSS images, identifying nearly 30,000 hot sds candidates. In addition, \citet{Zou2024} developed an MK-like spectral classification system for $\sim$1,200 hot sds using LAMOST~DR9 spectra, categorizing stars based on helium content and spectral characteristics. Meanwhile, \citet{Tahir2024} employed machine learning methods, specifically kernel support vector machines (SVM), to classify $\sim$11,000 B-type and $\sim$2,400 hot sds. Their analysis revealed that the linear kernel SVM achieved the highest accuracy, demonstrating the potential of spectrum-based classification to improve stellar identification. Similar techniques were also effective in detecting variability of hot sds. \citet{Ranaivomanana2025} used machine-learning algorithms to analyse the variability of $\sim$1,500 candidate hot sds using Gaia~DR3 and TESS data. Through dimensionality reduction techniques, the authors identify new hot sds variables, as well as potential new cataclysmic variables, demonstrating the effectiveness of this approach for classifying variable stars in large surveys. All these studies collectively highlight that these techniques have become indispensable in the new era of large-scale data analysis in astronomy. They emphasize the need to refine the techniques while identifying the most appropriate methods for each particular scientific case.

The most extensive catalogue of hot sds up to date was compiled by \citet{Culpan2022}, who identified $\sim$62,000 hot subluminous stars from Gaia~EDR3, representing a substantial improvement in detection in crowded Galactic regions and near the Magellanic Clouds. Building on this new catalogue, in this paper we will expand and further develop the classification techniques introduced in \citetalias{Viscasillas2024}. This approach aims to refine our understanding of hot sds populations, enhancing the analysis of their binary nature and astrophysical properties.\\

\par
This paper is structured as follows. Section~\ref{section: Data} describes our data set and the relation between XP~coefficients and stellar parameters. Section~\ref{section: Detection of binary hot sds systems based on XP coefficients} presents the detection of binary hot subdwarf systems using XP~coefficients. In Section~\ref{section: Detection of cool, He-poor hot subdwarfs (sdBs) based on normalised XP spectra}, we normalise the XP~spectra to emphasise individual spectral features and classify objects as cool/He-poor (sdB) or hot/He-rich. Section~\ref{section: Statistical analysis of the SOM and CNN training results} provides a statistical analysis of the SOM and CNN training results, while Section~\ref{section: Predicted labels for 20061 hot subdwarfs} presents the class predictions for all 20,061 objects. In the same section, we investigate correlations between binarity and stellar variability, including contamination by cataclysmic variable stars.

\section{Data}
\label{section: Data}

The basis of our sample is the list of $\sim$62,000 hot subdwarfs from \cite{Culpan2022}. There, the targets were selected by combining multi-band photometry and astrometric measurements from Gaia~EDR3. For a subset of their total hot hot subdwarf sample, \cite{Culpan2022} provide spectroscopically determined effective temperatures (T\textsubscript{eff}) and helium abundances (log(Y)). These parameters were compiled from the literature.

As Gaia’s most recent data release \citep[DR3;]{Gaiacol2023} has made abundantly clear, Gaia is not only a photometric and astrometric mission, but also a spectroscopic one. Its Radial Velocity Spectrometer \citep[RVS;]{Recio-Blanco2023} provides medium-resolution (R$\sim$11,500) spectra in the near-infrared Ca II triplet region (845–872 nm), while its photometric instrument includes two slitless prisms, the Blue and Red Photometers (BP and RP), that disperse light over the ranges 330$–$680~nm and 640$–$1050~nm, respectively \citep{Andrae2023}. These BP and RP data, collectively referred to as low resolution (R$\sim$50) XP spectra, were obtained simultaneously with the astrometric and photometric observations for an unprecedented number of stars, ensuring an exceptional level of internal consistency and calibration. Gaia DR3 released about 220 million mean XP spectra and 1 million mean RVS spectra, together with astrophysical parameters derived for hundreds of millions of sources. The agreement of Gaia’s XP-based parameters with ground-based surveys such as LAMOST \citep{Zhang2023} and of its RVS results with Gaia-ESO \citep{VanderSwaelmen2024} demonstrates the reliability of Gaia’s spectroscopic data, extending the mission’s impact well beyond astrometry.

In this work, we combine Gaia photometric, astrometric and low-resolution spectroscopic information. Although Gaia’s RVS spectra offer higher resolution than the XP data, only a small fraction (29 sources) of the $\sim$62,000 hot sds from \citet{Culpan2022} have RVS spectra available. This very limited overlap precludes the use of RVS data for any statistically meaningful analysis. We therefore rely exclusively on Gaia XP spectra, which provide homogeneous spectroscopic coverage for a substantial part of the sample. Our full data set consists of XP~coefficients of 20061 hot sds which are listed in \cite{Culpan2022}. These XP~coefficients are the coefficients of 110 basis functions, which in combination represent the full XP~spectra (in standard flux vs. wavelength format, \citealt{Carrasco2021}). The spectra cover the wavelength range from 330 to 1050~nm, in bins~of~2~nm. In Sections~\ref{subsection: Similarity map of Gaia XP coefficients} and \ref{section: Detection of binary hot sds systems based on XP coefficients} we use the 110 XP~coefficients to construct a similarity map of our objects and to detect binary systems in our sample. Following \cite{PerezCouto2024}, we normalise the coefficients of every hot sds by dividing them by their L2 norm\footnote{$L2(x) = \sqrt{x_1^2 + x_2^2 + ... + x_n^2}$, for a vector x with n components.}. The L2 norm, also known as Euclidean norm, is a measure for the length of a vector. Normalising the training data ensures that every input feature is treated with equal importance during the training process. Normalisation also increases the numerical stability and improves convergence of the training optimisation process \citep{Kim2025}. In Section~\ref{section: Detection of cool, He-poor hot subdwarfs (sdBs) based on normalised XP spectra}, we use the spectra in their flux-wavelength format.

\subsection{Similarity map of Gaia XP coefficients}
\label{subsection: Similarity map of Gaia XP coefficients}

\begin{figure}
  \centering
  \includegraphics[width=\columnwidth]{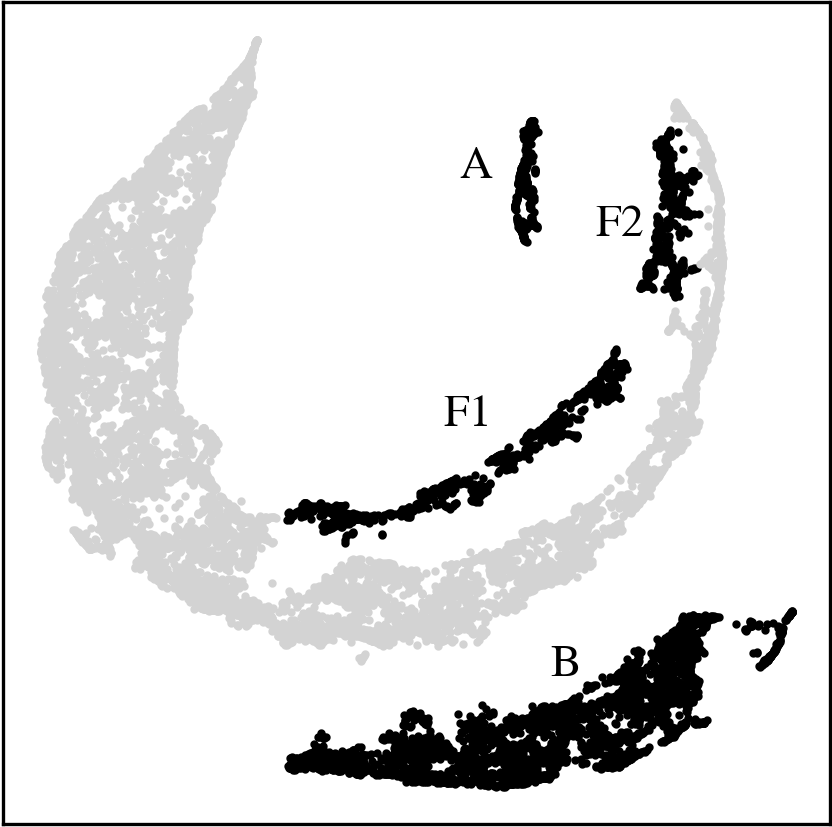}
  \caption{UMAP of the 20061 sets of XP coefficients. The main body of the map is coloured in grey, separate regions are kept in black and labelled A, B, F1, and F2.}
  \label{figure: UMAP_coeffs_groups}
\end{figure}

In a first step, we analyse the similarities and differences between the XP coefficients of our sample hot sds by using the Uniform Manifold Approximation and Projection (UMAP, \citealt{mcinnes2018}). The UMAP method is designed to reduce high-dimensional data sets to lower dimensions while keeping the internal relations between data points intact. In our case, the high-dimensional data are the 20061 sets of XP coefficients with 110 dimensions each (one dimension for every coefficient). To visualize this data, we reduce this set down to two dimensions with UMAP. The resulting similarity map is shown in Fig.~\ref{figure: UMAP_coeffs_groups}. In this map, every set of XP~coefficients is represented by one point, and the distance between two points depends on the similarity between the represented coefficient sets. The UMAP algorithm places more similar sets closer together, while different sets are placed further apart in the map. We see that our map is not uniform, but separated into different regions. Apart from the main body, which contains most of the stars in our sample, there are two detached islands (A and B) and two filaments F1 and F2.

\begin{sidewaysfigure*}
  \centering
  \makebox[0pt][c]{%
    \includegraphics[width=0.9\textheight]{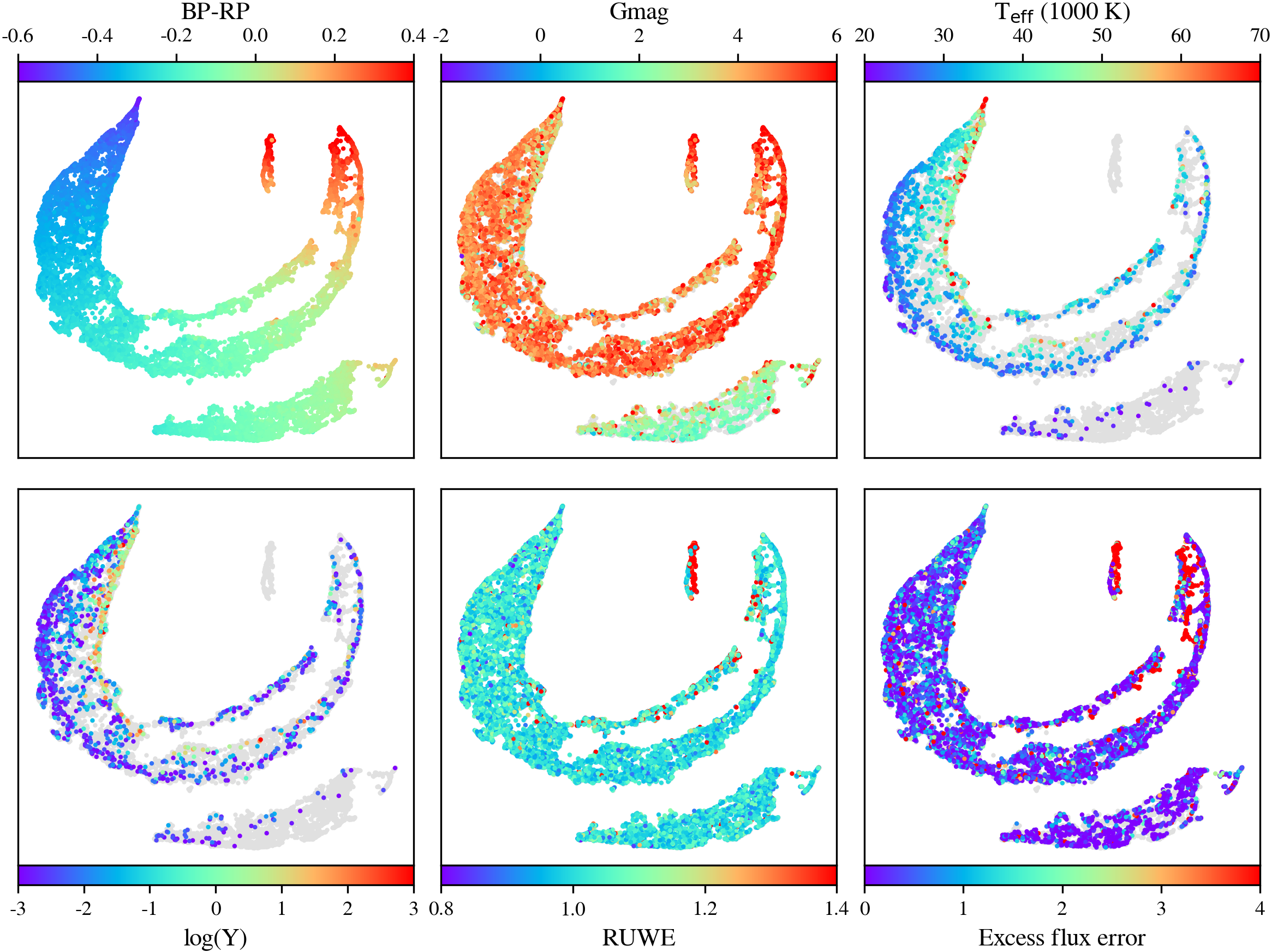}
  }
  \caption{UMAP of the 20061 sets of XP coefficients. Every panel shows the same UMAP, colour-coded by different parameters. Gray points represent spectra for which the respective parameter is not available.}
  \label{figure: UMAP_coeffs_parameters}
\end{sidewaysfigure*}

To investigate the physical differences between the stars in these regions, we colour-code the similarity map by six different stellar parameters from \cite{Culpan2022}. The six panels in Fig.~\ref{figure: UMAP_coeffs_parameters} show the same map with added colour dimensions. Colour-coding the points reveals correlations between certain areas of the map and different stellar parameters. This is expected since the spectra, and therefore the XP~coefficients of an observed star depend on the star's physical properties. Colour-coding by Gaia~BP-RP magnitudes, for example, shows that the similarity map is dominated by the colour-index: The coefficients of the bluest stars on the left side of the similarity map are most different from those of the red stars at the right edge of the map. A smooth colour gradient connects stars in those two extremes. This property is also apparent when we colour-code the processed spectra themselves by the Gaia BP-RP magnitudes (Fig.~\ref{figure: spectra_BP-RP}). Stars with red colours show a higher flux towards longer wavelengths than blue stars. As a consequence of this behaviour, the four UMAP groups A, B, F1, and F2 occupy different areas of the colour-magnitude diagram (Figure~\ref{figure: color-magnitude_umap-groups}). 

The distribution of absolute Gaia G~magnitudes is more uniform across the whole similarity map. However, the island B, at the bottom of the map, is brighter than the remaining regions. The average G~magnitude of island~B is 2.65~mag, approximately 2~mag brighter than the average of the other regions (4.68~mag).

The panels in the middle row of Fig.~\ref{figure: UMAP_coeffs_parameters} show that the UMAP method can separate the XP-coefficients of hot, He-rich stars from the cooler ones with lower He abundance. This correlation between stellar effective temperature and atmospheric He abundance in hot subdwarf stars has been observed before (for example, \citealt{Nemeth2012}). In our data set, the correlation coefficient between T\textsubscript{eff} and log(Y) is 0.51.

The bottom row of Fig.~\ref{figure: UMAP_coeffs_parameters} shows the colour-coding according to the renormalised unit weight error (RUWE) and excess flux error. The RUWE parameter is a dimensionless measure for how much the centre of an object moves between different observations, \citep{Lindgren2021} and is therefore a measure of astrometric variability. The excess flux error \citep{Gentile_Fusillo2021} is an indicator for the photometric variability of a source. 

\clearpage
\clearpage

Large values of the excess error indicate an unusually high scatter of measured brightness between different observations, when compared to the brightness scatter of other objects with similar absolute magnitude, colour, and number of observations. A high value of the excess flux error therefore suggests that the measured brightness variations are intrinsic to the object itself, and not resulting from observational errors. The excess flux errors used in this study have been calculated by \cite{Culpan2022} for about 60\% of the hot sds in our sample. In Section~\ref{subsection: Connection between indicators of stellar variability and binarity}, we investigate the connection between the RUWE and excess flux errors, as indicators of stellar variability, and the hot sds binarity in our sample.

Almost all stars that have high values for both RUWE and the excess flux error are found in the island A of our similarity map. The island A itself is split between a half that shows indications of variability (right) and an invariable half (left). The filament F2 contains most of the remaining hot sds with large photometric variability.

\subsection{Island A: contamination by objects that are not hot subdwarfs}
\label{subsection: Island A: contamination by objects that are not hot subdwarfs}

None of the objects in island A have previously been studied in \cite{Solano22} or in \citetalias{Viscasillas2024}. Furthermore, none of the references cited in \cite{Culpan2022} report any T\textsubscript{eff} or log(Y) values for these stars.
To better understand the nature of the island A objects, we analysed several of them with the VOSA tool. Out of the total 561 objects in island A, 58 satisfy the criteria for a reliable VOSA fit. These criteria demand the presence of photometric data in both GALEX (Galaxy Evolution Explorer; \citealt{GALEX2005}) FUV and NUV filters, as well as coverage in the 3D extinction maps provided by \cite{Lallement2022}.

The spectral energy distributions of all 58 objects can be fitted with models for cool main sequence stars, specifically BT-Settl models with T\textsubscript{eff} ranging from 4000 to 10000~K \citep{Allard2012}. Four of these objects can be well fitted by a single BT-Settl model, meaning that they are most likely single stars. The remaining 54 require a combination of two models to achieve a good fit and are therefore classified as binaries by VOSA. All 58 objects are too cool to be hot subdwarfs, which typically have effective temperatures exceeding 19000~K. Examples of VOSA fits of a typical binary hot sds and a cooler sample object from island A are shown in Fig.~\ref{figure: VOSA_fits}.

\begin{figure}
  \centering
  \includegraphics[width=\columnwidth]{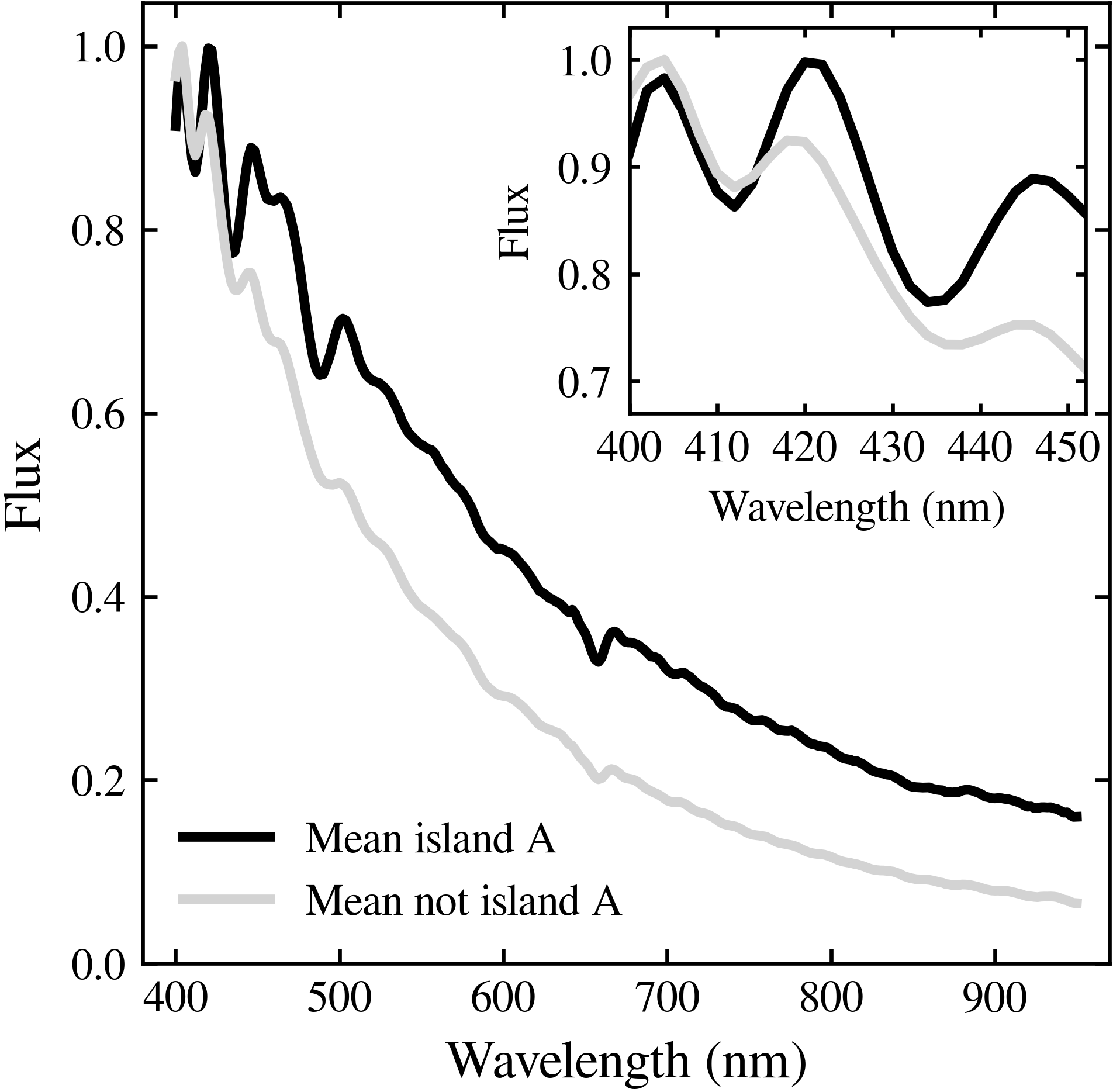}
  \caption{Average of Gaia XP~spectra of all objects in island A (black) and average XP~spectrum of all other samples that are not is island A (grey). The insert panel highlights the region from 400 to 550~nm.}
  \label{figure: spectra_island_A}
\end{figure}

We also conducted a visual inspection of the SEDs of the island A objects. While confirmed hot subdwarfs exhibit a characteristic negative slope in the UV region, the island A objects display a positive slope. Figure~\ref{figure: spectra_island_A} presents the average XP~spectra of the island A objects alongside the average spectra of the remaining sample. The most striking difference is the elevated flux in the island A spectra, which is approximately 0.1 units higher than the mean flux of the other spectra. This flux difference remains nearly constant at wavelengths beyond 500~nm.

In the range of $\sim$400 to 420~nm, the continuum of the mean island A spectrum remains flat (the absorption feature at 410~nm is a hydrogen Balmer line). In contrast, the mean flux of the non-island A spectra decreases by $\sim$0.1 over the same 20~nm range. This flux drop is the typical negative slope in the blue part of the spectrum seen in hot objects (for example, hot subdwarfs). 

The shallower slope at short wavelengths, combined with the increased flux at longer wavelengths, suggests that the island A objects are cooler than the remaining sample.

This inspection of the XP~spectra supports the VOSA analysis of the 54 binary candidates in island A, confirming that they are too cool to be hot subdwarfs. We see in the colour-magnidute diagram in Fig.~\ref{figure: color-magnitude_umap-groups}, that island A contains many of the reddest objects in our sample. They lie close to the cutoff line which was defined by \cite{Culpan2022} to separate the hot subdwarf candidates from the main sequence. This, together with the low temparatures, makes it very likely that most of the island A objects are simply main sequence stars. However, a detailed investigation into the nature of island A lies beyond the scope of this work.

This analysis demonstrates that similarity maps are a powerful tool for identifying objects that are physically distinct from the main population of our sample. In our reported results (Section~\ref{section: Predicted labels for 20061 hot subdwarfs}), we include a flag to identify the island A objects.

\section{Detection of binary hot sds systems based on XP coefficients}
\label{section: Detection of binary hot sds systems based on XP coefficients}

When we compare the binary distribution in the map to the top left panel of Fig.~\ref{figure: UMAP_coeffs_parameters}, we see that the main difference between the filament binaries and the non-filament binaries is the BP-RP colour index. For 29 of the VOSA binaries, properties of their companion star have been measured by \cite{Solano22}. This allows us to investigate how the filament binaries differ from the non-filament ones. Of the 29 binaries with companion information, 22 are in one of the two filaments. The remaining seven objects are scattered across the rest of the map. The companion effective temperatures and radii are not significantly different in the two groups. However, the filament binary companions are more luminous. We performed a Mann-Whitney U test to decide if the difference between the luminosity distributions of the two groups is statistically significant. This test is suitable for small samples, for which it is not clear if they are normally distributed. The Mann-Whitney U p-value for our samples is $\sim$0.01. This is smaller than the critical p-value of 0.05, which confirms the difference between the two groups. The mean bolometric luminosity (L\textsubscript{bol}) of the filament companions is equal to 0.44~L$_{\odot}$ and they contribute on average $\sim$3.3\% to the total luminosity of the binary system. The non-filament companions have a mean L\textsubscript{bol} of 0.18~L$_{\odot}$, which amounts to less than 1\% of the total system flux, on average. Figure~\ref{figure: UMAP_coeffs_Lbol_companions} shows the contribution of the companion to the total bolometric luminosity of the binary system across our similarity map. For $\sim$131 of our filament binaries, \cite{Culpan2022} report the spectral class of the companion star. All except one of the filament companions are main sequence (MS) stars, either labelled as hot subdwarf~+MS, or more specifically, +F, +G, or +K. The remaining companion is a white dwarf (hot sds~+WD). Main sequence companions are also found outside the filaments. However, most of the non-filament companions (64\%) are either white dwarfs or low mass M-dwarfs (hot~sds+dM). Three of the non-bilament companions are brown dwarfs (sds~+BD). These faint companions do not significantly contribute to the total luminosity, and therefore to the spectral energy distribution of their binary systems. This suggests that the binaries located on the left-hand side of the UMAP diagram are systems in which the contribution of the secondary component is very small, which explains the overlap and confusion between single and binary hot subdwarfs in this region.
For a more detailed classification of our sample into single and binary hot subdwarf systems, we train use two machine learning methods SOM and CNN. The class predictions by both methods for our full sample of 20061 objects is presented in Section~\ref{section: Predicted labels for 20061 hot subdwarfs}.

\begin{figure}
  \centering
  \includegraphics[width=\columnwidth]{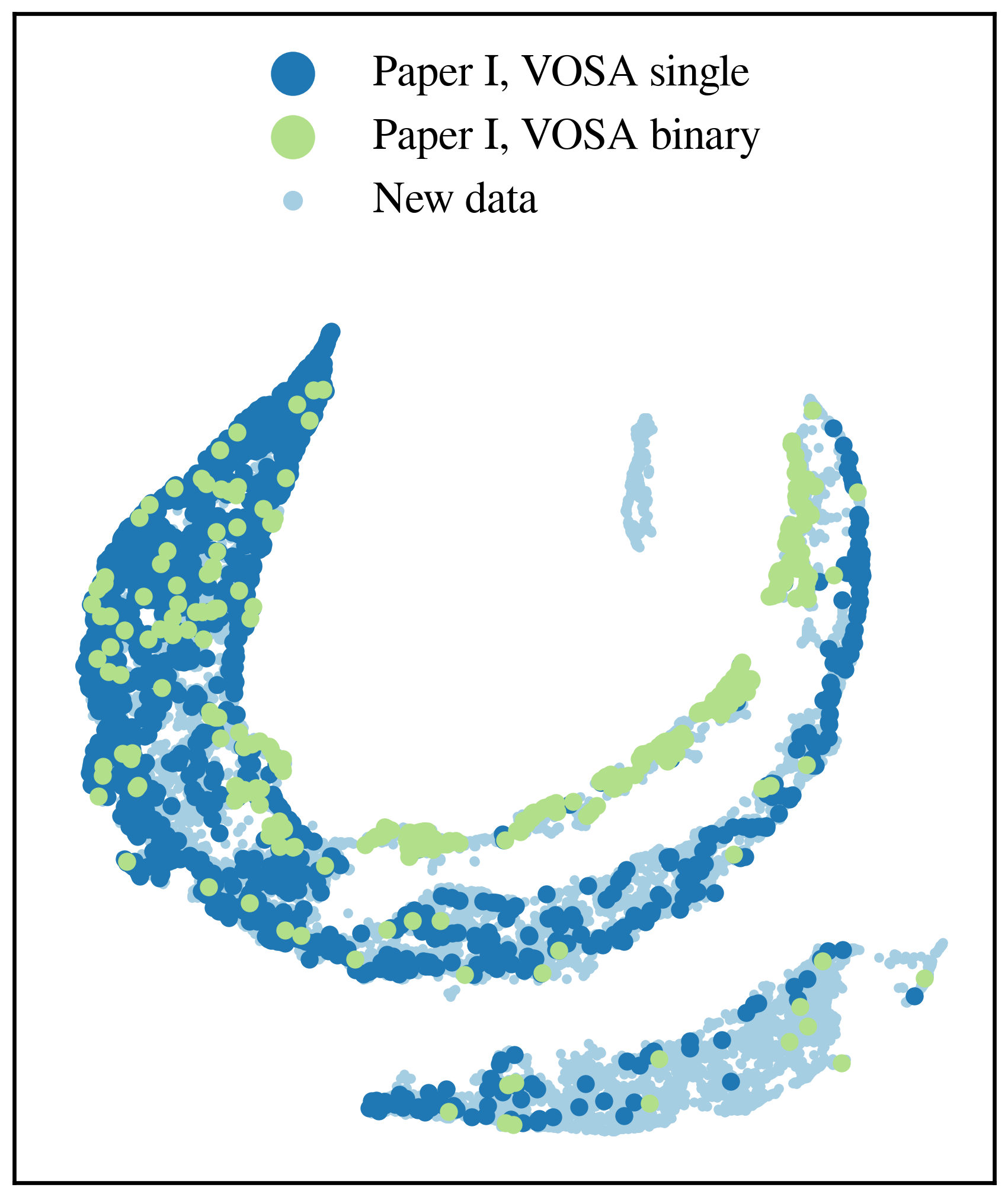}
  \caption{Same UMAP as in Fig.~\ref{figure: UMAP_coeffs_groups}, colour-coded by VOSA binarity. Our new data set has 2695 samples in common with set from \citetalias{Viscasillas2024}.}
  \label{figure: UMAP_coeffs_VOSA}
\end{figure} 

\subsection{Binary detection with SOM}
\label{subsection: Binary detection with SOM}

We leveraged the SOM algorithm to further explore the different populations of our dataset. The main advantage of the SOM with respect to other unsupervised machine learning techniques is that the SOM is both a dimensionality reduction algorithm and a clustering technique. Indeed, the SOM projects a high-dimensional non-linear dataset (such as the XP coefficients of our sources) in a 2D map, preserving the topology order of the feature space, such as the UMAP. 
However, in the SOM, the 2D map is a finite grid of neurons so that similar elements fall in the same neuron, and neurons with similar elements are placed adjacent to each other. In this manner, we can perform statistics and study each neuron separately, assigning labels (classes) to each neuron according to the predominant class from the labeled data that fall into it.
To implement it, we used the \texttt{MiniSom}\footnote{\url{https://github.com/JustGlowing/minisom/}} \citep{vettigliminisom}. After some hyperparameter tuning we chose a $8 \times 8$ map size, $\sigma = 1.4$, learning rate $0.5$, and the Euclidean distance as metric. The number of iterations was $5000$.
To classify a neuron $z_{(i,j)}$ (and thus the sources contained inside it) with a label $X$ we require the neuron to contain at least 10 labeled sources to have enough statistical power. Any neuron that does not meet that condition is considered as an outlier, and will not be used for classification.

\begin{figure}
    \centering
    \includegraphics[width=\columnwidth
    ]{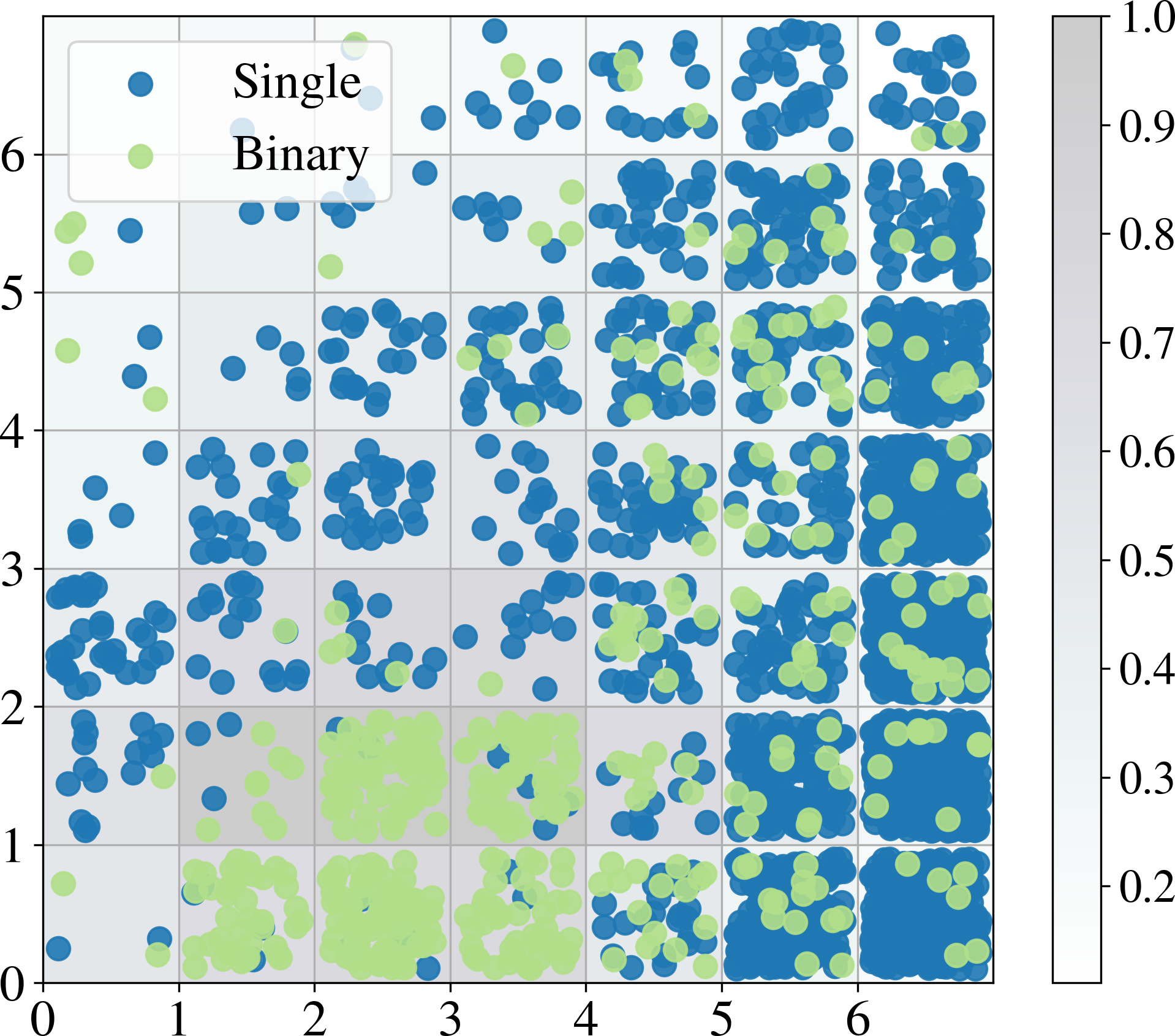}
    \caption{SOM map of the XP coefficients, with known hot sds colour-coded according to their binarity.}
    \label{figure: SOM_binarity}
\end{figure}

We apply the SOM to the XP coefficients of our sample of $\sim20,000$ hot sds. As a result, we present in Fig.~\ref{figure: SOM_binarity} the SOM map with colour-coded binarity. We used the training set of $\sim2700$ hot sds with single/binary flag from VOSA for the colours. Sources with unknown binarity are plotted with transparent colour for better visualization. As can be seen, there are six neurons in the lower left part of the SOM with an evident representation ($P_\text{binary}>0.90$) of binary hot sds, encompassing a total of 302 hot sds binary candidates, while the rest of the known hds binaries are uniformly mixed with single hot sds. Eleven neurons are classified as outliers, and therefore, 38 neurons are finally used for the classification.

\subsection{Binary detection with CNN}
\label{subsection: Binary detection with CNN}

Before training our CNN, we created a test set by randomly selecting 1000 of the total 2695 training samples. This test set is not involved in the network training process. It is instead used to assess how well the trained CNN performs on previously unseen data. For the training, the remaining 1695 samples will be split into a validation set of 500 samples and the preliminary training set of 1195 samples. We train our CNN until its performance on the validation set does not increase any more. This prevents our CNN from overfitting to the training data. We improve upon our classification model from \citetalias{Viscasillas2024} by addressing the single-binary class imbalance in the training set and by fine-tuning the network training process.

Our preliminary training set of 1195 samples is imbalanced, with only 17\% belonging to the binary class. To counter this imbalance during the network training, we apply two different methods. First, we supplement our training set with synthetic binary samples using the Synthetic Minority Over-sampling Technique (SMOTE, \citealt{imbalanced-learn}). Using SMOTE, we generate artificial sets of binary samples by interpolating between existing ones. With this technique, we roughly double the number of binaries in our training set to a final binary ratio of 33\%. Adding even more synthetic binaries to the training set worsens the performance of the trained CNN on the test set. The final training set then contains 1477 samples, with 492 belonging to the binary class and 985 to the single class. We do not add any artificial samples to the validation and test sets. To counter the remaining class imbalance in the training set, we apply different loss weight factors to the binary and single samples (0.66 for binary samples, 0.33 for single samples). In this way, the network will be penalized more when it misclassifies binaries. This causes the network to pay more attention to the binary samples.

Due to the small number of samples in our training set, combined with the uneven distribution of binaries in the similarity map, it is possible that random sampling splits our data into non-uniform training and validation sets. For example, if a large majority of the selected training set binaries is located in the filament F1 of the similarity map, then the CNN cannot learn to accurately predict labels for the main body of the map. To minimize the effects of this possible imbalance, we trained an ensemble of ten CNNs. We create new validation and training sets before each of the training runs. After all CNNs have been trained, we inspect their performances on the test set. We keep the five CNNs with the highest binary accuracy scores and discard the remaining five. The binary probabilities that we provide in Section~\ref{section: Predicted labels for 20061 hot subdwarfs} are the average predicted values from the best five networks.

\section{Detection of cool, He-poor hot subdwarfs (sdBs) based on normalised XP spectra}
\label{section: Detection of cool, He-poor hot subdwarfs (sdBs) based on normalised XP spectra}

\begin{figure*}[]
    \centering
    \includegraphics[width=\the\textwidth]{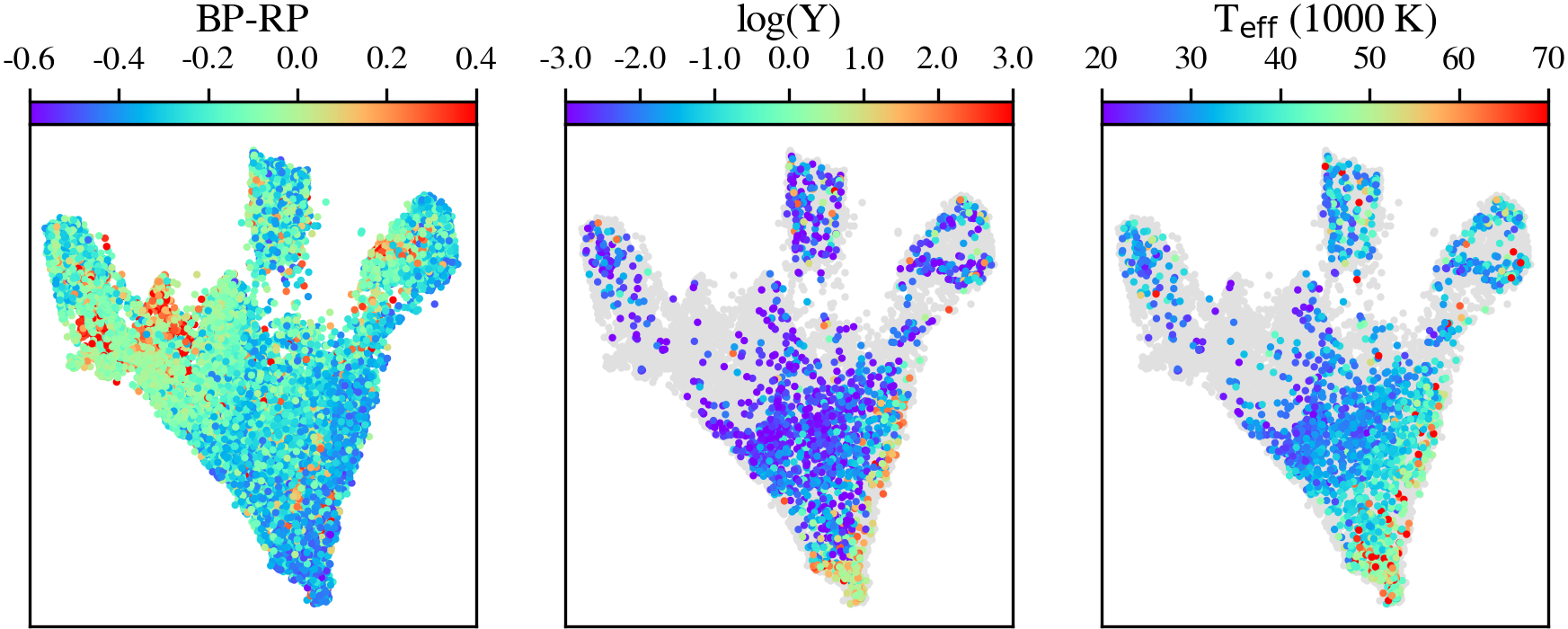} \caption{UMAP of 20023 normalised XP spectra. All panels show the same UMAP, colour-coded by different parameters. Grey points represent spectra for which the respective parameter is not available.}
    \label{figure: UMAP_norm_params}
\end{figure*}

In Section~\ref{subsection: Similarity map of Gaia XP coefficients}, we observed that the position of a hot sds in the coefficient similarity map primarily depends on its BP-RP colour. The same is true when we generate a similarity map for the XP~spectra of our sample. This is because the overall shape of a spectrum is dominated by its flux distribution at longer wavelengths, and therefore by the colour of the star. The more subtle features of our spectra are overwhelmed by this dominance of the overall flux distribution. Therefore, the individual features play a less significant role in the grouping of stars in the similarity map.
To remove the influence of the intrinsic spectral flux distribution, we normalise our spectra to the continuum. We determine the continuum baselines using the asymmetric least squares smoothing (ALS) method \cite{Eilers&Boelens2005}. By dividing each spectrum by its continuum baseline, we obtain the normalised spectra.
We repeat the same UMAP analysis as in Section~\ref{subsection: Similarity map of Gaia XP coefficients}, this time with the normalised XP~spectra. The resulting similarity map is shown in Fig.~\ref{figure: UMAP_norm_params}. The colour-coding confirms that the normalization has reduced the BP-RP colour's influence on the overall shape of the spectra. This allows individual absorption features to become more prominent in the total spectral shape. The middle panel of Fig.~\ref{figure: UMAP_norm_params} shows that the He-rich hot sds are grouped closer together in this similarity map than they are in the map of the raw XP~coefficients. We also observe that the normalised XP~spectra of high-temperature hot sds occupy the same space in the map as the He-rich objects. This indicates that in our data set both of these parameters have the same effect on the normalised spectra. 

\begin{figure}
    \centering
    \includegraphics[width=\columnwidth]{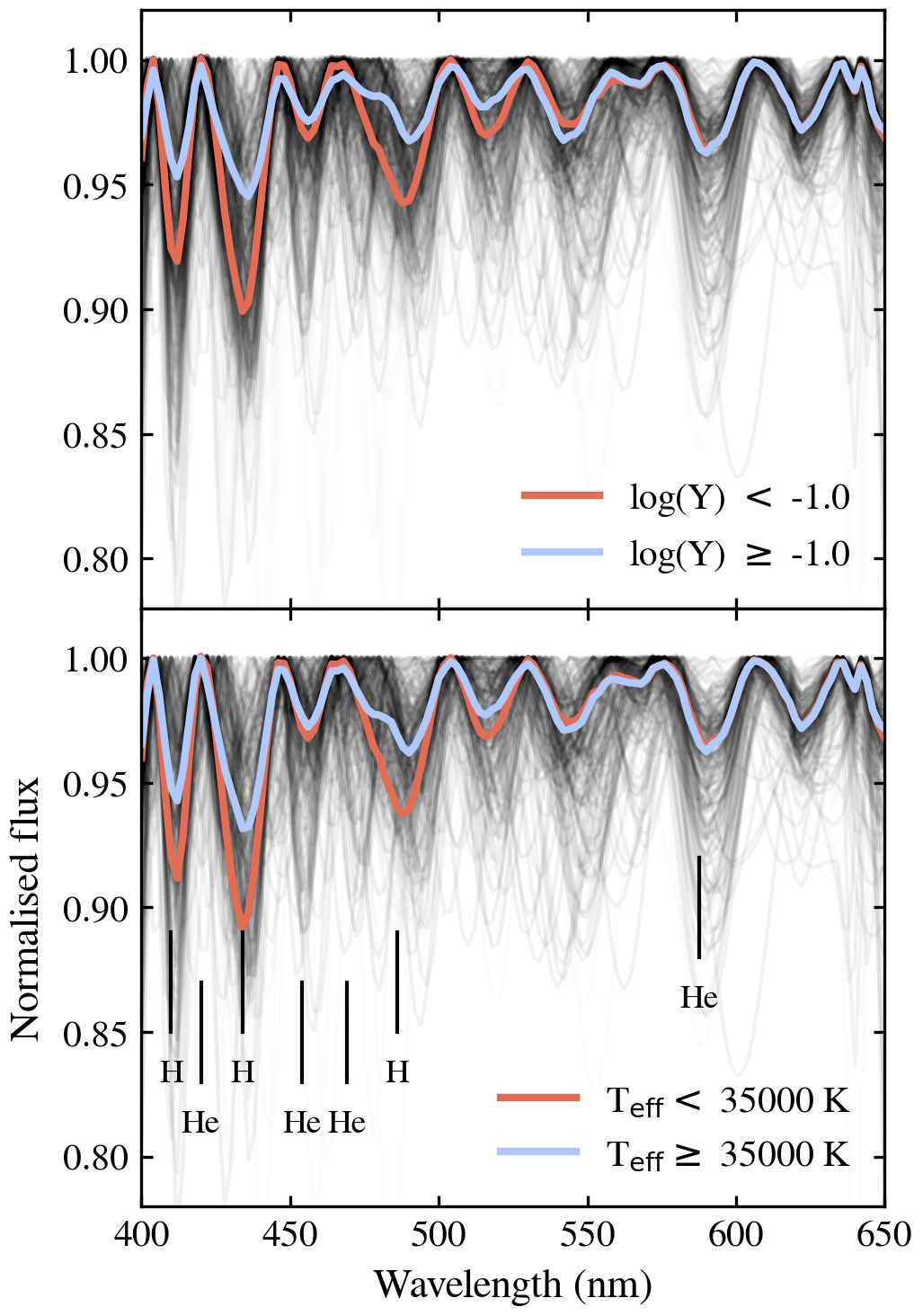}
    \caption{Top panel: Mean normalised spectra of He-poor (red) and He-rich (blue) hot sds. Bottom panel: Mean normalised spectra of cool (red) and hot (blue) hot sds. The locations of several relevant absorption lines are indicated in the bottom panel. Their wavelengths are 410, 434, and 486~nm (HI); 420, 454, 469~nm (HeII), and 588~nm (HeI). In both panels, the mean spectra are plotted together with 300 randomly selected normalised spectra from our full sample (black).}
    \label{figure: spectra_norm_He_Teff}
\end{figure}

To see the range of spectral shapes for samples with different He abundances, we plot the mean normalised spectra of all He-rich and He-poor stars in the top panel of Fig.~\ref{figure: spectra_norm_He_Teff}. We chose the threshold between He-rich and He-poor hot sds to be log(Y)~=~-1. The choice of this threshold value is motivated by the He abundance classification scheme for hot sds by \cite{Luo2021}. We see that the two categories differ most strongly at the short wavelength part of the spectral range. The He-poor sample shows stronger absorption features than the He-rich sample. This is counter-intuitive because more helium in a stellar atmosphere is expected to absorb more light and therefore cause stronger spectral absorption features. The wavelengths with the largest differences between mean He-rich and He-poor spectra are 410, 434, and 486~nm. These are not the locations of He absorption lines, but the wavelengths of the hydrogen Balmer lines H$\delta$, H$\gamma$, and H$\beta$. At the low resoltuion of the XP~spectra, the weaker He-absorption features are fully blended with the much stronger hydrogen absorption lines. In the bottom panel of Fig.~\ref{figure: spectra_norm_He_Teff} we observe that the effective temperature has the same effect on the spectral shape as the He abundance. The hydrogen Balmer lines of the hottest stars, with T$_\mathrm{eff}$~>~35000~K, are weaker than those in the cooler spectra. With rising T$_\mathrm{eff}$, a larger fraction of the hydrogen atoms in the stellar atmospheres becomes ionised and can therefore no longer absorb. This results in weaker H-absorption features in hotter stars. We can see from this comparison of both panels in Fig.~\ref{figure: spectra_norm_He_Teff} that our XP~spectra do not allow us to disentangle effects of the He-abundance from those of the effective temperature. 
Therefore, we separate our sample spectra into a sample that is both hot and He-rich (weak absorption features), and a cooler He-poor class (strong absorption features). More than 90\% of the samples in the cool/He-poor group are classified as sdB stars in \cite{Culpan2022}. We therefore label the samples in this class as "sdB". We define sdB samples as those objects with T$_\mathrm{eff}$ < 35000~K and log(Y) < -1. The hotter and more He-rich group is labelled as "hot/He-rich".The majority of stars in this class belong to one of the hot sds subtypes sdO, He-sdO, or He-sdB.
As in Section~\ref{section: Detection of binary hot sds systems based on XP coefficients}, we train our SOM and CNN tools to classify our sample objects into sdB stars and hot/He-rich sds. The classification is based on the normalised XP~spectra. The results for the whole sample are presented in Section~\ref{section: Predicted labels for 20061 hot subdwarfs}.

\subsection{Detection of sdB stars with SOM}
\label{subsection: Detection of sdB stars with SOM}

\begin{figure}
    \centering
    \includegraphics[width=\columnwidth]{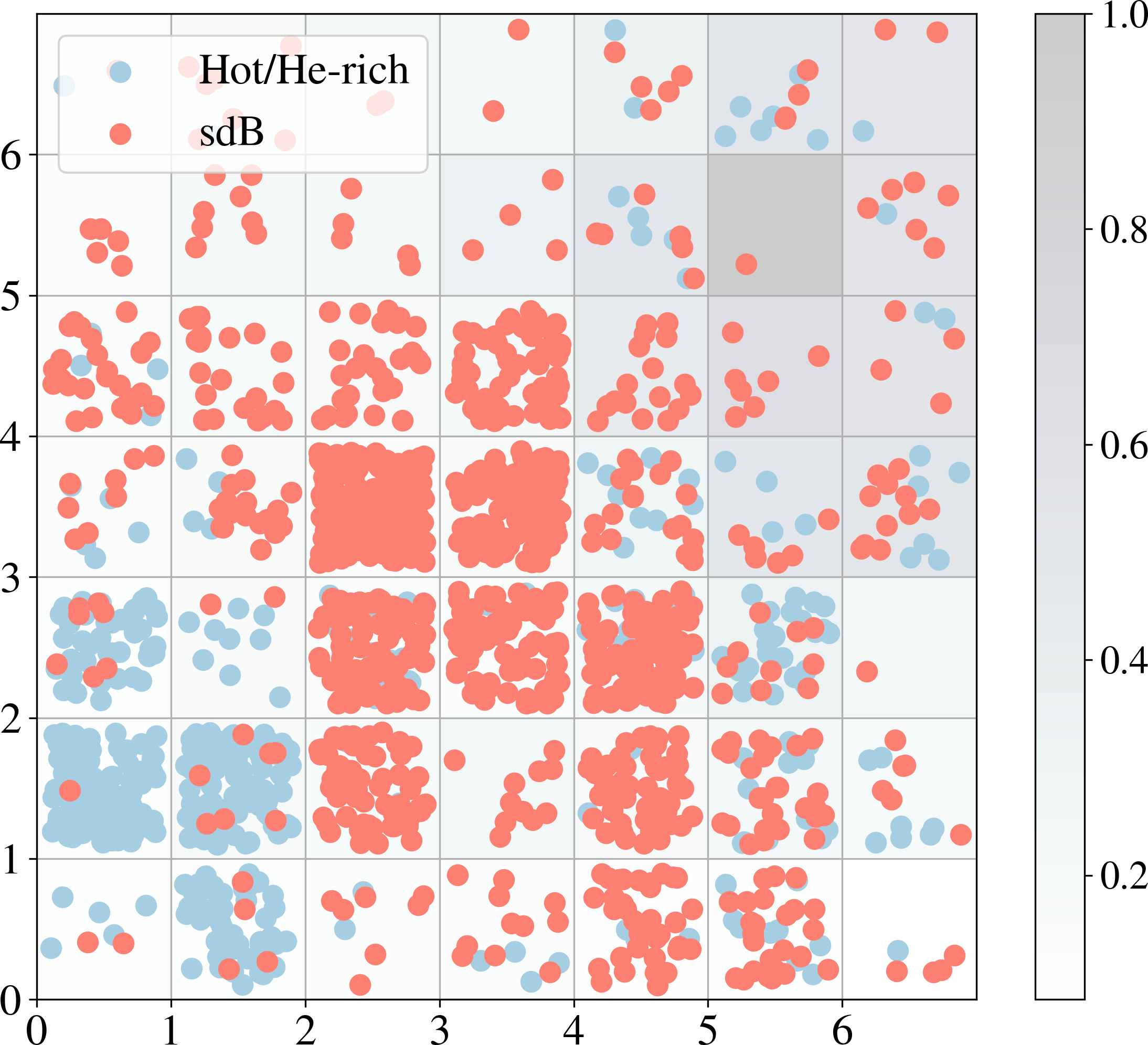}
    \caption{SOM map of the XP spectra, normalised to the continuum, with known hot sds colour-coded according to if they are He-rich or He-poor hot sds.}
    \label{figure: SOM_sdB}
\end{figure}

To use the SOM to select between hot/He-rich and sdB spectra, we repeat the steps in the previous section but using the spectra normalised to the continuum instead of the XP~coefficients. As done with the UMAP, the aim of this is to put the focus in the absorption lines instead of the continuum shape. We also normalise the input vector by the $\mathcal{L}_2$ norm and use the $\sim 2000$ sources from \citet{Culpan2022} with available T\textsubscript{eff} and $\log{Y}$ to colour-code the resulting SOM with hot/He-rich and sdB labels. The result is presented in Fig.~\ref{figure: SOM_sdB} and shows that there are two neurons, $z_{(0,1)}$ and $z_{(1,0)}$ (row 0 and column 1, and row 1 and column 0, respectively), with $P_{hot/He-rich} > 0.5$, while 15 are sdB neurons, and the remaining 32 are outliers.

\subsection{Detection of sdB stars with CNN}
\label{subsection: Detection of sdB stars with CNN}

Analogous to the search for binary hot sds in Sect.~\ref{subsection: Binary detection with CNN}, we now use CNNs to classify the normalised XP spectra into sdB and hot/He-rich. Of our total 20074 sample spectra, 1959 have both T$_\mathrm{eff}$ and log(Y) available in the data of \cite{Culpan2022}. Of these, 250 spectra were assigned to each the test set and validation set. The training set then contains 1459 spectra. The fraction of sdB spectra in the training set is 54\%, the rest is classified as hot/He-rich. As for the binary detection task, we train an ensemble of ten CNNs and keep the five with the best performance on the test set. The reported results are the average predictions from the five best CNNs. Since our sdB and hot/He-rich classes are nearly balanced, we do not apply SMOTE sampling or class loss weights in the training phase.

Network gradients show which parts of the input spectra have the most influence on the CNN output. Figure~\ref{figure: CNN_gradients_sdB} shows the network gradients for the class sdB, together with average spectra of the two classes in our training set. The gradients have the deepest negative peaks at the positions of the hydrogen absorption lines. Negative sdB gradients indicate that a decrease of the flux (that is, deeper absorption) around these wavelengths increases the CNN probability for the class sdB. We already observed in Fig.~\ref{figure: spectra_norm_He_Teff} that the difference between our two classes is largest at the hydrogen absorption lines. The network gradients therefore show that our CNN can separate sdB from hot/He-rich samples in a physically meaningful and intuitive way.

\begin{figure}
    \centering
    \includegraphics[width=\columnwidth]{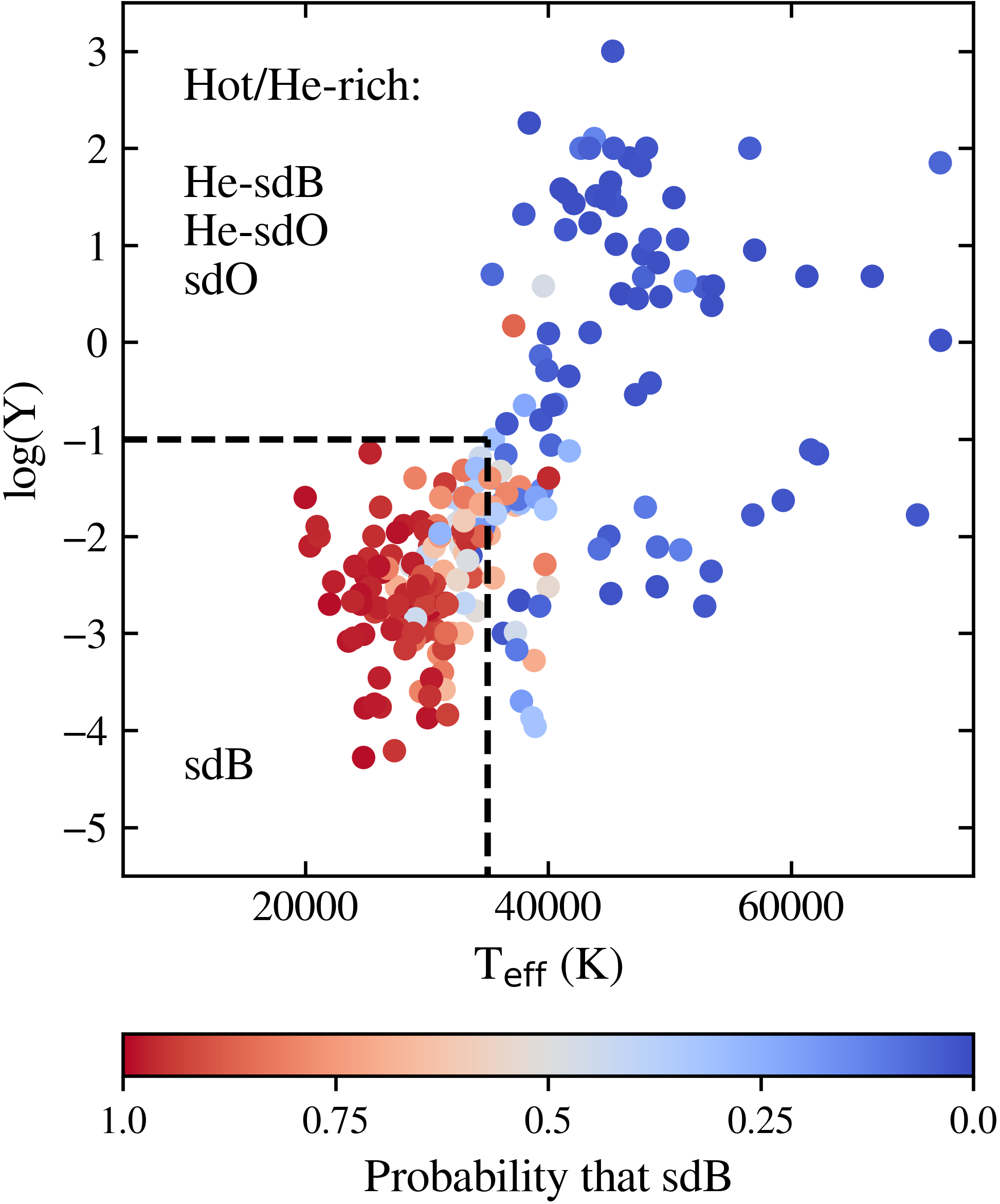}
    \caption{Helium abundances log(Y) relative to the effective temperatures T$_\mathrm{eff}$ of the stars in the test set. The colour scale shows the probability of belonging to the sdB class (cool/He-poor), as predicted by our CNN ensemble. The dashed lines show our defined class boundaries of T$_\mathrm{eff}$~=~35000~K and log(Y) = $-$1.0~dex.}
    \label{figure: CNN_prediction_sdB_test}
\end{figure}

As the training result, we show the CNN predictions for the test set in Fig.~\ref{figure: CNN_prediction_sdB_test}. We see that the CNN ensemble predicts high probabilities for belonging to the class sdB for stars that have temperatures and helium abundances within our set thresholds. Close to the class boundaries, the network classification is less certain for several spectra. The probabilities for these samples is closer to 50\%. However, the cooler, He-poor stars are clearly not classified as hot/He-rich by our network ensemble. Given the correlation between T$_\mathrm{eff}$ and He-abundance in O and B stars, we have no cool stars with high He-abundances in our sample (neither in the test set nor in the training or validation sets). Therefore, the separation between our two classes appears to be based on the temperature alone. Still, as discussed above, we cannot distinguish between the effects of temperature and He-abundance in our set of XP spectra.

\section{Statistical analysis of the SOM and CNN training results}
\label{section: Statistical analysis of the SOM and CNN training results}

Since both CNN and SOM models are probabilistic binary classifiers, there are three main metrics one has to use when evaluating model performance. These are sensitivity (TPR), specificity (TNR) and precision (PPV), which are defined as follows

\begin{equation}
    \begin{aligned}
    \text{TPR}=\frac{\text{\# of true positives}}{\text{\# of true positives}+\text{\# of false negatives}},\\
    \text{TNR}=\frac{\text{\# of true negatives}}{\text{\# of true negatives}+\text{\# of false positives}},\\
    \text{PPV}=\frac{\text{\# of true positives}}{\text{\# of true positives}+\text{\# of false positives}}.
    \end{aligned} 
\end{equation}  

\subsection{ROC and PR curves}
\label{subsection:ROC and PR curves}

Examining the interaction between TPR, TNR and PPV provides a fundamental means of assessing the model’s performance. We begin our analysis by plotting the Receiver Operating Characteristic (ROC) and Precision–Recall (PR) curves and calculating the corresponding Area Under the Curve (AUC) scores for both models and datasets (see Fig. \ref{figure: ROC_curves}, Fig. \ref{figure: PR_curves}, and Table \ref{table: scores}). These visualizations illustrate the trade-offs between TPR and TNR (in the ROC curve) and between TPR and PPV (in the PR curve) across different decision threshold values. A higher AUC value, closer to 1, indicates better overall model performance.

As we observe, the CNN model demonstrates strong performance, with ROC AUC values around 0.95 and PR AUC values around 0.9. In contrast, the SOM model performs considerably worse, particularly on the hot/He-rich stars dataset. For both models, classification performance is lower on the hot/He-rich stars dataset compared to the binary star dataset.
It is also worth noting the Log Loss value of 1.051 for the SOM model on the hot/He-rich stars dataset, indicating that the model is overly confident when making incorrect predictions. Furthermore, we observe pronounced diagonal segments in the SOM model’s ROC curve for the hot/He-rich stars dataset (see the bottom right panel in Fig.~\ref{figure: ROC_curves}). This indicates that the SOM model cannot effectively distinguish positives from negatives in those regions (i.e., locally random performance).

\begin{table}[!ht]
\caption{Performance evaluation scores.}
\centering
\begin{tabular}{lccc}
\toprule\toprule
     Score & Model  & Binary stars & hot/He-rich stars \\
\midrule
ROC AUC & SOM  & 0.906   & 0.839   \\
ROC AUC & CNN  & 0.961   & 0.946   \\
PR AUC & SOM  & 0.793   & 0.838   \\
PR AUC & CNN  & 0.915   & 0.943   \\
Brier & SOM  & 0.064   & 0.187   \\
Brier & CNN  & 0.039   & 0.089   \\
Log loss & SOM  & 0.244   & 1.051   \\
Log loss & CNN  & 0.160   & 0.294   \\
\bottomrule
\end{tabular}
\label{table: scores}
\end{table}

\subsection{Threshold optimisation}
\label{subsection: Threshold optimisation}

Next, we aim to identify the threshold values that yield the best performance for our models. However, this optimization depends on how “good” performance is defined. To optimize specificity, we employ Youden’s index, defined as

\begin{equation}
\text{Youden’s index} = \text{TPR} + \text{TNR} - 1,
\end{equation}

while to optimize precision, we use the $F_1$ score, defined as

\begin{equation}
F_1 = \frac{2 \cdot \text{PPV} \cdot \text{TPR}}{\text{PPV} + \text{TPR}}.
\end{equation}

After threshold optimization (see Table~\ref{tab:thresholdCombined}), we note that all previously observed trends are maintained. That is, the CNN model consistently outperforms the SOM model in terms of sensitivity, specificity, and precision. Additionally, both models perform better on the binary hot sds dataset than on the hot/He-rich stars dataset.

\begin{table}[h!]
\caption{Optimal model threshold results for Youden's index and $F_1$ score.}
\centering
\resizebox{\columnwidth}{!}{
\begin{tabular}{llcccc}
\toprule\toprule
Model & Dataset & Threshold & Sensitivity & Specificity & Precision \\
\midrule
\noalign{\smallskip}
\multicolumn{6}{l}{\textbf{Youden's Index}} \\
\noalign{\smallskip}
SOM & Binary & 0.177 & 0.759 & 0.915 & 0.649 \\
SOM & hot/He-rich & 0.169 & 0.773 & 0.738 & 0.714 \\
CNN & Binary & 0.295 & 0.878 & 0.965 & 0.838 \\
CNN & hot/He-rich & 0.745 & 0.818 & 0.985 & 0.920 \\
\midrule
\noalign{\smallskip}
\multicolumn{6}{l}{\textbf{$F_1$ Score}} \\
\noalign{\smallskip}
SOM & Binary & 0.367 & 0.664 & 0.972 & 0.832 \\
SOM & hot/He-rich & 0.102 & 0.859 & 0.625 & 0.660 \\
CNN & Binary & 0.745 & 0.818 & 0.985 & 0.920 \\
CNN & hot/He-rich & 0.525 & 0.847 & 0.908 & 0.886 \\
\bottomrule
\end{tabular}
}
\label{tab:thresholdCombined}
\end{table}

\section{Predicted labels for 20061 hot subdwarfs}
\label{section: Predicted labels for 20061 hot subdwarfs}

In this section, we present the single vs. binary and hot/He-rich vs. sdB class predictions for our full sample of 20061 hot sds. We then use these results to analyse the correlation between stellar variability indicators and hot sds binarity. Table~\ref{table: results_CNN_SOM} lists the CNN and SOM label predictions for the first few stars in our sample. A full version of this table is available online.

We see in Section~\ref{section: Statistical analysis of the SOM and CNN training results}, that the CNN ensemble is more precise than the SOM in predicting the classes for the training sets. The following analysis is therefore based on the CNN predictions, using the probability threshold values that optimize the prediction precision (F1 score, see Table~\ref{tab:thresholdCombined}). \par
For the threshold value of P(binary)~=~0.745, our CNN ensemble predicts a binary ratio of $\sim$17\ in our full sample. This corresponds to 3359 detected binaries in our full data set. The found binary ratio in the full sample is the same as in our binarity training set. The CNN ensemble for the hot/He-rich vs. sdB classification predicts a ratio of $\sim$66\% sdB in the full sample (using the classification threshold 0.525). This corresponds do a number of 13174 sdB stars in our sample. \par
For our CNN ensemble, the reported class probabilities are the average values from the five best-performing models. Along with these averages, we also include the standard deviations (std) of the class probabilities across the five models. These deviations reflect the internal uncertainty of the CNN ensemble (see, for example, \citealt{Guiglion2024}). We find that this internal uncertainty is low when the predicted class probability is close to 0 or 1. In these cases, the model is confident, and the samples are clearly representative of their respective classes. However, as the predicted probability gets closer to 50\%, the uncertainty increases. Figure~\ref{figure: CNN_uncertainties} shows the internal uncertainty of the CNN ensemble relative to the predicted class probability for the binary and sdB classes.

\subsection{Connection between indicators of stellar variability and binarity}
\label{subsection: Connection between indicators of stellar variability and binarity}

In this section, we investigate the connection between hot sds binarity and the indicators for astrometric and photometric variability. Following \cite{Culpan2022}, we consider a hot sds as variable if its RUWE parameter is larger than 1.4 or Excess flux error > 4.0. These thresholds divide the RUWE vs. Excess flux error diagram into four regions: region 1 with no indication of variability, region 2 showing photometric variability only, region 3 showing both photometric and astrometric variability, and region 4 with only astrometric variability. 

Of our 20061 sample objects, 11639 have measurements of both RUWE and excess flux errors (excluding the island A objects). Figure~\ref{figure: variability_binarity} shows the distribution of this hot sds sample in the RUWE versus Excess flux error diagram. Of the 11639 hot sds, 90.5\% are invariable, 6.2\% show indications of photometric variability, only 0.6\% display both photometric and astrometric variability, and 2.7\% show photometric variability alone.
We find a correlation between stellar variability and binarity. Starting from a binary ratio of 17.6\% in the invariable region 1, the ratio increases in the other regions. Region 2 has a binary ratio of 60.5\%, region 3 shows 77.1\%, and region 4 has 80.8\% binaries.

\subsection{Contamination by cataclysmic variable stars}
\label{Contamination by cataclysmic variable stars}
\begin{figure}
  \centering
  \includegraphics[width=\the\columnwidth]{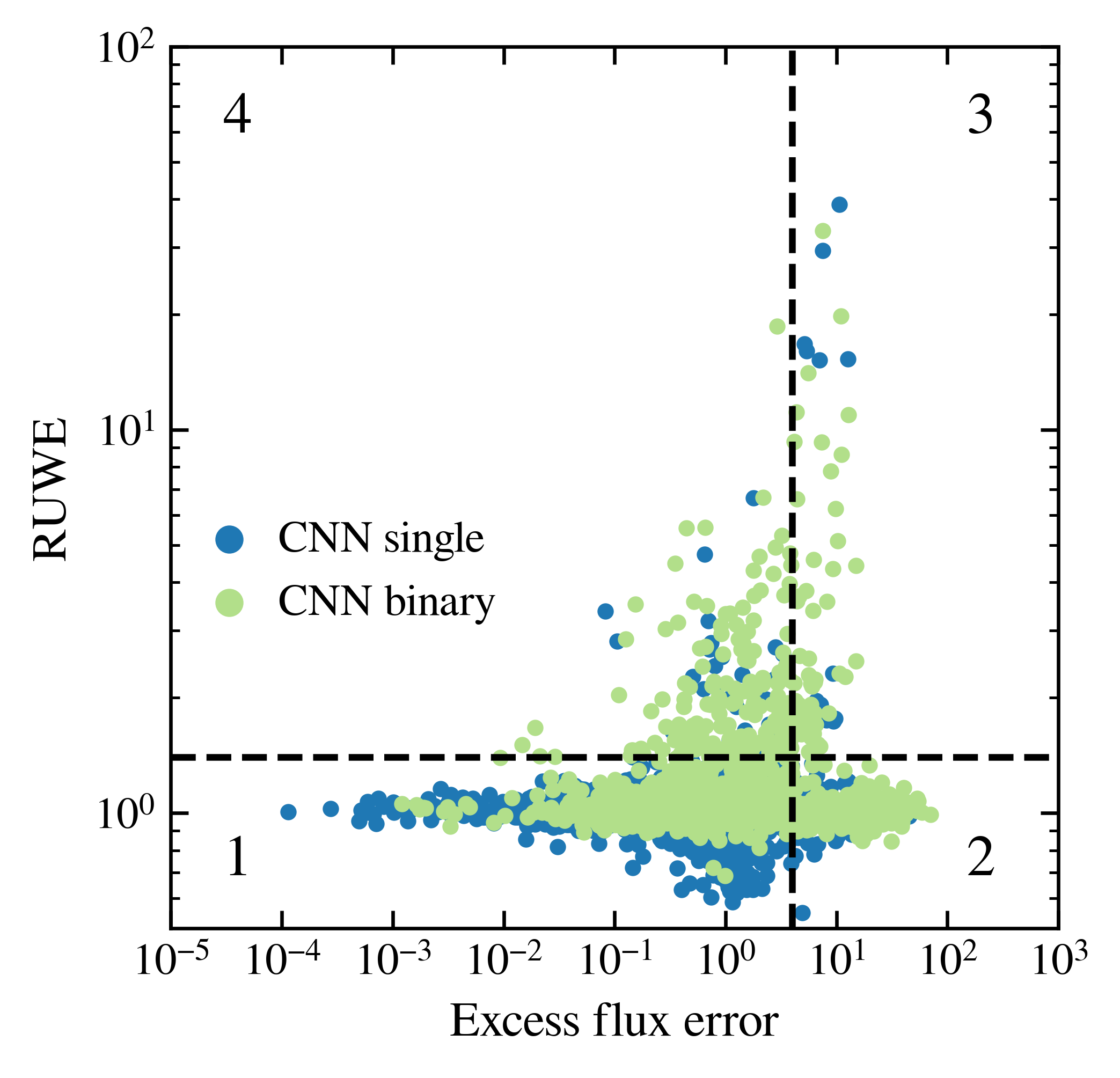}
  \caption{Gaia RUWE versus Excess flux error of hot sds in our sample. CNN single stars are coloured in blue, CNN binaries in green. Stars from the island A are not included in this plot. The dashed lines show the variability cutoffs at RUWE~=~1.4 and Excess flux error~=~4.0 used in \cite{Culpan2022}. The resulting regions of the diagram are numbered from 1 to 4.}
  \label{figure: variability_binarity}
\end{figure}

We remind the reader that the binaries that we detect in this work are those with composite spectra. These are generally wide systems with orbital periods of several hundred days. Wide orbits cause increased astrometric wobble, and therefore higher RUWE values \citep{Culpan2022}. This leads to the observed correlation between RUWE values and binary ratios in regions 3 and 4 in Fig.~\ref{figure: variability_binarity}. The elevated binary ratio in region 2, which only shows photometric variability, is harder to explain. The distance between hot subdwarf and wide-orbit companion is too large for the two components to interact directly, for example through the transfer of material from the hot sds to the companion. High photometric variability of our detected binary systems is therefore unexpected. A possible explanation is that the region 2 contains contamination by cataclysmic variables (CVs). These objects are binary systems, composed of a white dwarf that accretes material from its companion, typically a late-type star (e.g., \citealt{Smak2001}). Many CVs occupy a region in the Hertzsprung–Russell diagram that overlaps with the reddest objects in our sample (BP$-$RP~$\sim$0.5, GMAG~$\sim$5.0; \citealt{Abril2020}). To evaluate a possible contamination of our sample by CVs, we crossmatched our data set with several catalogues of cataclysmic variables. These are the Open Cataclysmic Variable Catalog \citep{Jackim2024}, the Gaia~DR3 catalogue of variable sources \citep{Eyer2023}, and the catalogue of cataclysmic variables of \cite{Inight2023}. In total, we find 203 unique CV sources in our full sample, and all except five of them are located in the region 2 of our Fig.~\ref{figure: variability_binarity}. The 198 CVs in the region 2 make up $~$28\% of the 720 objects in this variability region. Many cataclysmic variables are known X-ray sources, contributing a large part of the total galactic X-ray emission. To find additional contamination with CVs that were not listed in the above catalogues, we crossmatched our sample with the first data release catalogue of the X-ray survey eROSITA \citep{Merloni2024}. We find a match of 120 X-ray sources in our full sample, 67 of them not in any of the CV catalogues. Of these 67 X-ray sources, 45 show high photometric variability. Interestingly, the majority of the found CVs in our sample are concentrated in the filament 2 in our UMAP of XP~coefficients (region F2 in Fig.\ref{figure: UMAP_coeffs_groups}). This finding suggests that it is possible to search for cataclysmic variables by using Gaia-XP spectra. While a deeper investigation of the CV contamination in our sample is beyond the scope of this current work, it is clear that a large fraction of our detected binaries with high excess flux errors are not hot sds binaries, but couuld be cataclysmic variables. Our binarity probability predictions for samples with RUWE < 1.4 and excess flux error > 4 should therefore be used with caution.

\section{Summary and conclusions}
\label{section: Summary and conclusions}

Taking the list of $\sim$62,000 hot subdwarfs from \cite{Culpan2022} as our basis sample, a selection of targets combining multi-band photometry and astrometric Gaia measurements, together with available physical parameter values, a subsample of 20061 hot subdwarf stars was retained for study. Several techniques were employed, with a main aim focused on analysing their possible binary properties and classification.

In order to gain an overview over our data set, we used UMAP to reduce the high-dimensional Gaia XP~coefficients of 20,061 stars to two dimensions. The resulting similarity map contains several distinct groups of objects with similar XP coefficients. Colour-coding the map by various stellar parameters reveals the connections between the physical properties of our sample objects and their XP~coefficients. We find that stellar BP-RP colour is the dominant parameter that influences the position of an object in the similarity map. To a lesser extent, stellar effective temperatures, helium abundances, and stellar variability indicators are also important parameters that influence the location of a sample in our similarity map. The majority of VOSA classified binary star systems are grouped together in two filamentary structures in the similarity map. The filament companions are exclusively main sequence stars, that contribute on average $\sim$3.3\% to the total L\textsubscript{bol} of their binary systems. Fainter companions like low mass M-dwarfs, brown dwarfs, and white dwarfs are not found in the filaments, but are scattered across the rest of the similarity map. The UMAP dimensionality reduction reveals a group of 561 objects whose XP~coefficients are different from the coefficients of the rest of our sample. The investigation of 58 members of this separate group with the VOSA tool reveals that these objects are too cool to be either single or binary hot sds. A visual inspection of the SEDs and Gaia XP-spectra of these objects further shows the absence of the negative UV slope, which is typical for hot objects. Based on these indications of lower stellar temperatures, we conclude that the 561 objects in island A are not hot sds.

To find binary systems among our full sample, we trained the machine learning methods SOM and CNN. These tools can predict the classes binary or single for every hot sds, based on its XP-coefficients. The training set for both SOM and CNN consists of of a subset of 2695 hot sds, for which VOSA labels were available. For the supervised method CNN, we had to address the class imbalance in the training set, where only 17\% are binary class samples. By using SMOTE, we generated synthetic binary samples to increase the binary ratio to 33\%. In total, we trained an ensemble of ten CNN models, and used the best five models to predict the binary probability for our sample of 20061 hot sds. \par
To remove the influence of the flux excess, we normalise the XP spectra of our sample hot sds to the continuum. We determine the continuum baselines using the asymmetric least squares smoothing (ALS) method. The similarity analysis of the normalised spectra confirms that the BP-RP colour’s influence on the spectra shape has been significantly reduced, allowing individual absorption features to become more prominent. The spectra of hot and He-rich hot sds are grouped much closer together in the similarity map of the normalised spectra, compared to the coefficient similarity map. \par
At the low resolution of the Gaia XP spectra, the He absorption features are blended with the strong Balmer absorption lines of hydrogen. As a consequence, we cannot distinguish between the influence of temperature and the influence of He abundance on the shape of our spectra. We can therefore only separate the cool/He-poor spectra (equivalent to the hot sds subtype sdB) from the hot/He-rich spectra (He-sdB, He-sdO, and sdO). Analogous to the detection of binaries from XP coefficients, we trained SOM and an ensemble of CNNs to classify our normalised spectra into sdB and hot/He-rich samples.\par
After training the SOM and CNN ensemble, we use them to predict the binarity and sdB vs. hot/He-rich classifications for our full sample of 20061 objects. Our binarity predictions from the CNN ensemble show that there exists a correlation between astrometric and photometric variability indicators and hot sds binarity. The binary ratio among hot sds without indication of astrometric or photometric variability is $\sim$18\%, and increases to >60\% for stars with only photometric variability. We find that at least 28\% of the objects with only photometric variability are not hot sds, but instead  could be cataclysmic variable binary systems. Binarity probability predictions for samples with RUWE < 1.4 and excess flux error > 4 should therefore be used with caution. The binary ratio rises to $\sim$77\% for hot sds with both astrometric and photometric variability. At $\sim$82\%, the binary ratio is highest for hot sds that show only astrometric variability. \\
\par
As the next step of this work, we have recently initiated high-resolution spectroscopic observations with the Vilnius University Echelle Spectrograph \citep[VUES;][]{Jurgenson2016} for selected targets in our sample. This will allow us to validate our results and explore their binary properties in greater depth.

\section{Data availability}
\label{section: Availability of data and code}

The XP~coefficients and XP~spectra of our sample are available for download at \url{https://github.com/mambr-astro/Hot_sds_GaiaXP}. Python codes for the construction of the coefficient similarity map, the normalisation of XP~spectra, and the training of CNN ensambles for binary and sdB star detection are also available there.

\begin{acknowledgements}
We sincerely thank the anonymous referee for her/his valuable guidelines and insightful comments, which have significantly enhanced the quality of this work.
The authors acknowledge support from the project "Unveiling the Nature of Hot Subdwarfs through Spectroscopy and Machine Learning" (MSF-JM-17/2025), funded by the Vilnius University Research Promotion Fund. We also acknowledge support from EU HORIZON-CL4-2023-SPACE-01-71 SPACIOUS project, Ref. 101135205; the Spanish Ministry of Science MCIN / AEI / 10.13039 / 501100011033 and the EU FEDER collaborate through the coordinated grant PID2024-157964OB-C22. We also acknowledge support from the Xunta de Galicia and the European Union ED431B 2024/21, CITIC ED431G 2023/01. Ana Ulla acknowledges partial funding, under grant IFC3002/25, from the Fundación Ceo, Ciencia e Cultura. This research has made use of the Spanish Virtual Observatory (https://svo.cab.inta-csic.es) project funded by MICIU/AEI/10.13039/501100011033 and by ERDF/EU through Grant PID2023-146210NB-I00.
\end{acknowledgements}

\bibliographystyle{aa}
\bibliography{bibliography}

\begin{appendix}
\section{Complementary figures and tables}

\setlength{\textfloatsep}{10pt plus 2pt minus 2pt}  \setlength{\floatsep}{8pt plus 2pt minus 2pt}       \setlength{\intextsep}{8pt plus 2pt minus 2pt}      

\begin{figure}[!ht]
  \centering
  \includegraphics[width=0.7\columnwidth]{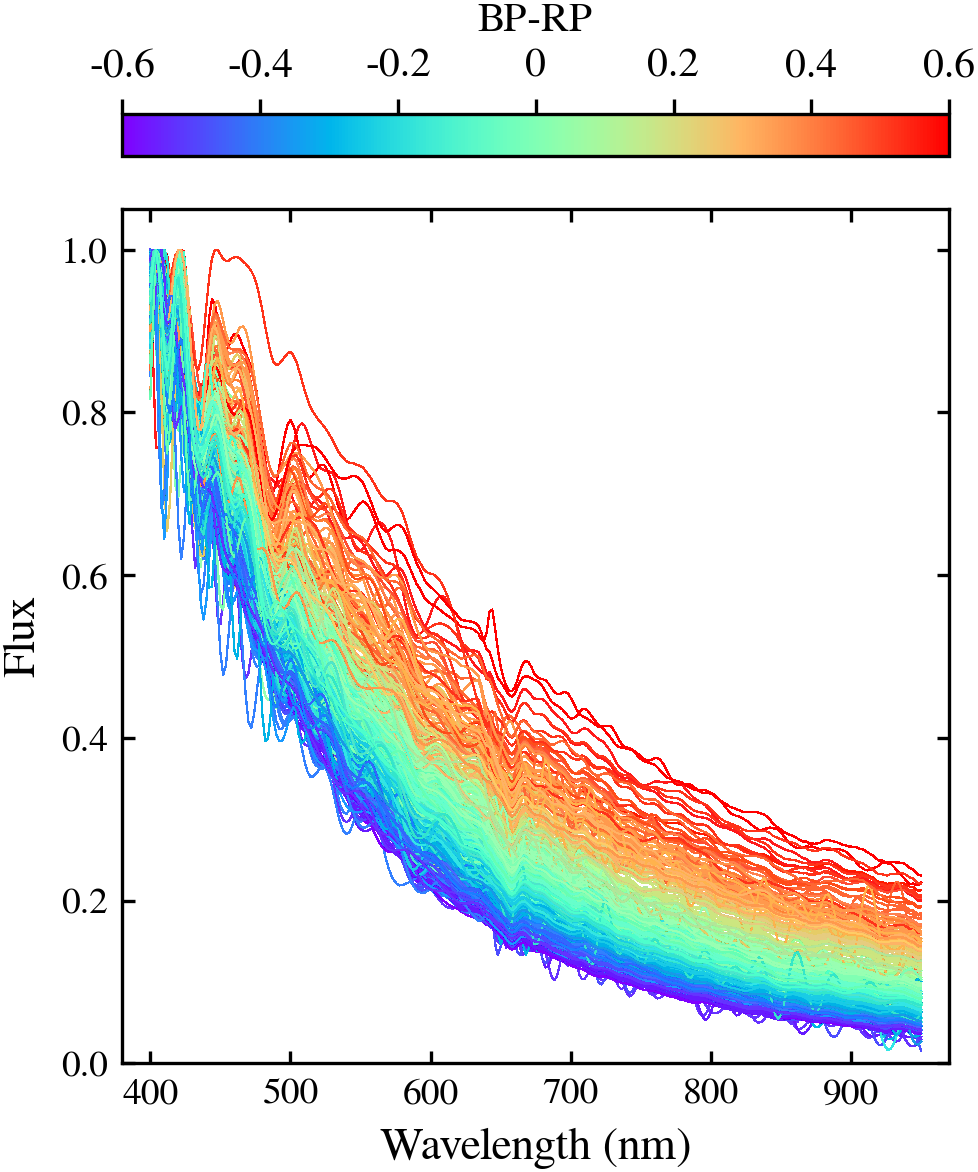}
  \caption{Gaia XP spectra of 750 randomly selected stars from our sample, colour-coded by Gaia BP-RP magnitudes. The spectra have been scaled to a maximum flux value of 1.0 and truncated to a wavelength range between 400 and 950~nm.}
  \label{figure: spectra_BP-RP}
\end{figure}

\begin{figure}[!ht]
  \centering
  \includegraphics[width=0.67\columnwidth]{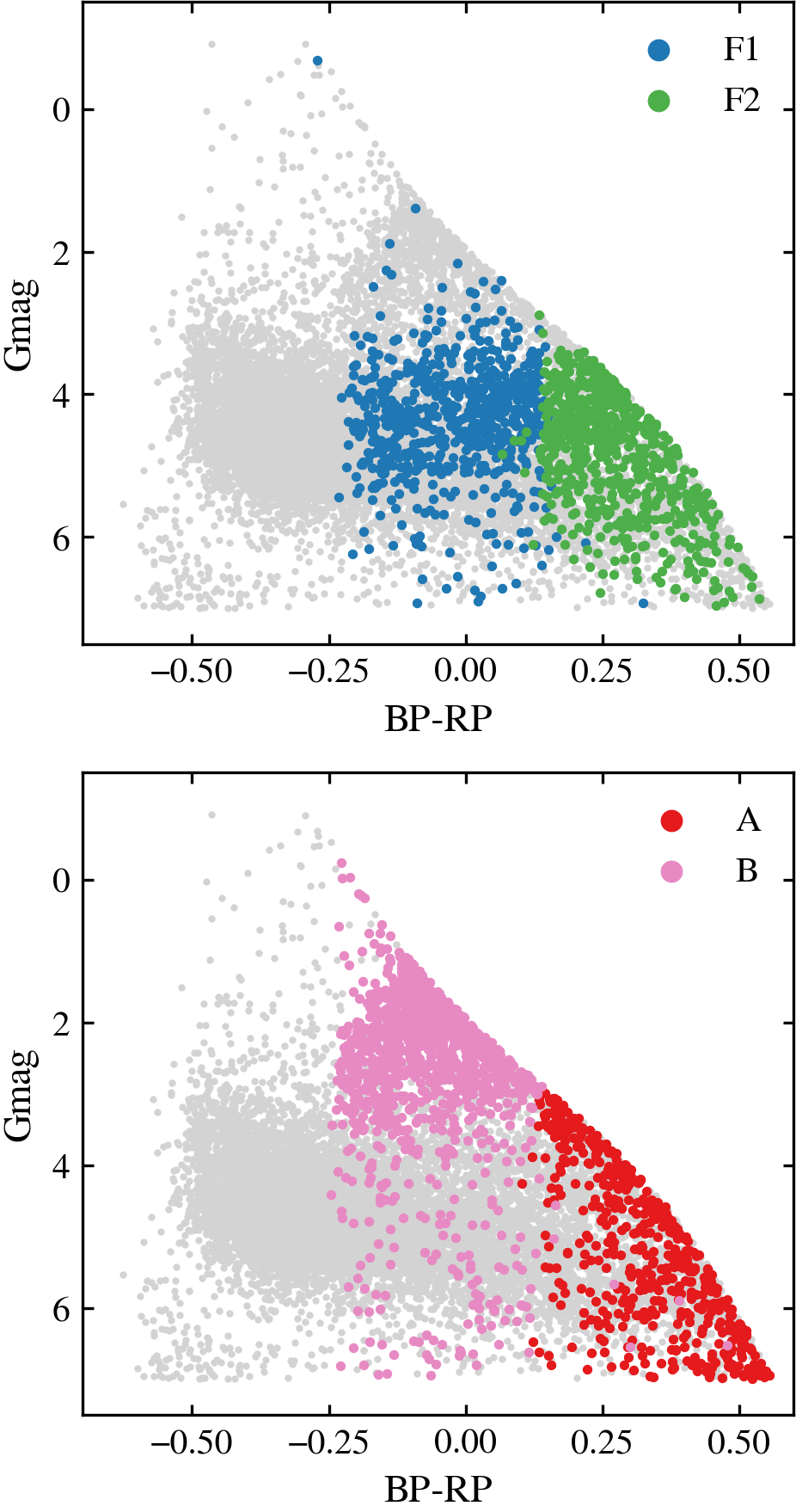}
  \caption{Colour-magnitude diagram of our sample objects, based on Gaia~DR3 photometry. Top panel: The UMAP filament regions are highlighted. Bottom panel: The UMAP island regions are highlighted. In both panels, the main body of the map is coloured in grey.}
  \label{figure: color-magnitude_umap-groups}
\end{figure}

\begin{figure}[!ht]
  \centering
  \includegraphics[width=0.9\columnwidth]{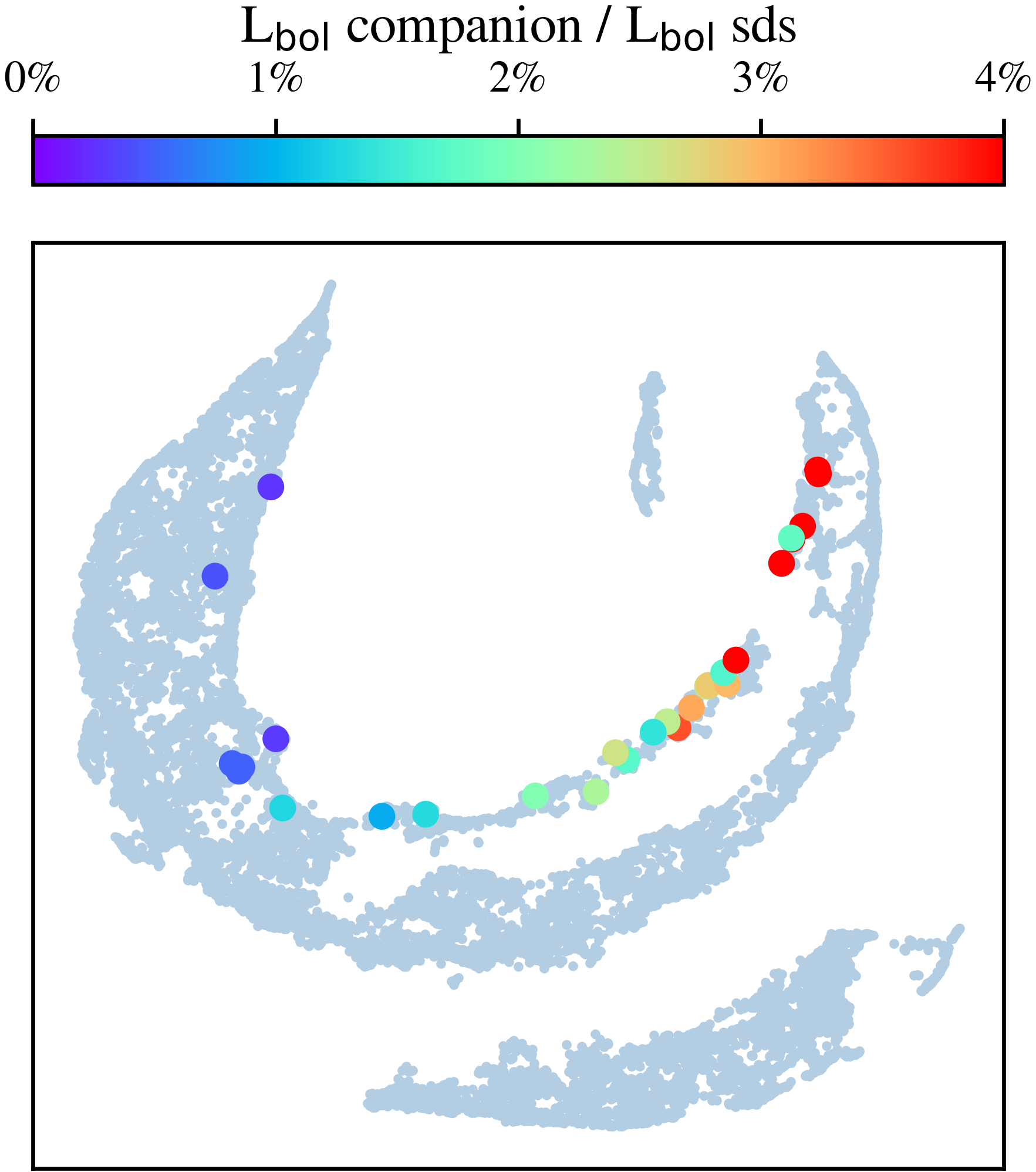}
  \caption{Similarity map of our 20061 sets of XP coefficients. Where available, colours indicate the ratio of the binary luminosity of the binary companion relative to the hot sds luminosity.}
  \label{figure: UMAP_coeffs_Lbol_companions}
\end{figure}

\begin{figure}[!ht]
  \centering
  \includegraphics[width=1\columnwidth]{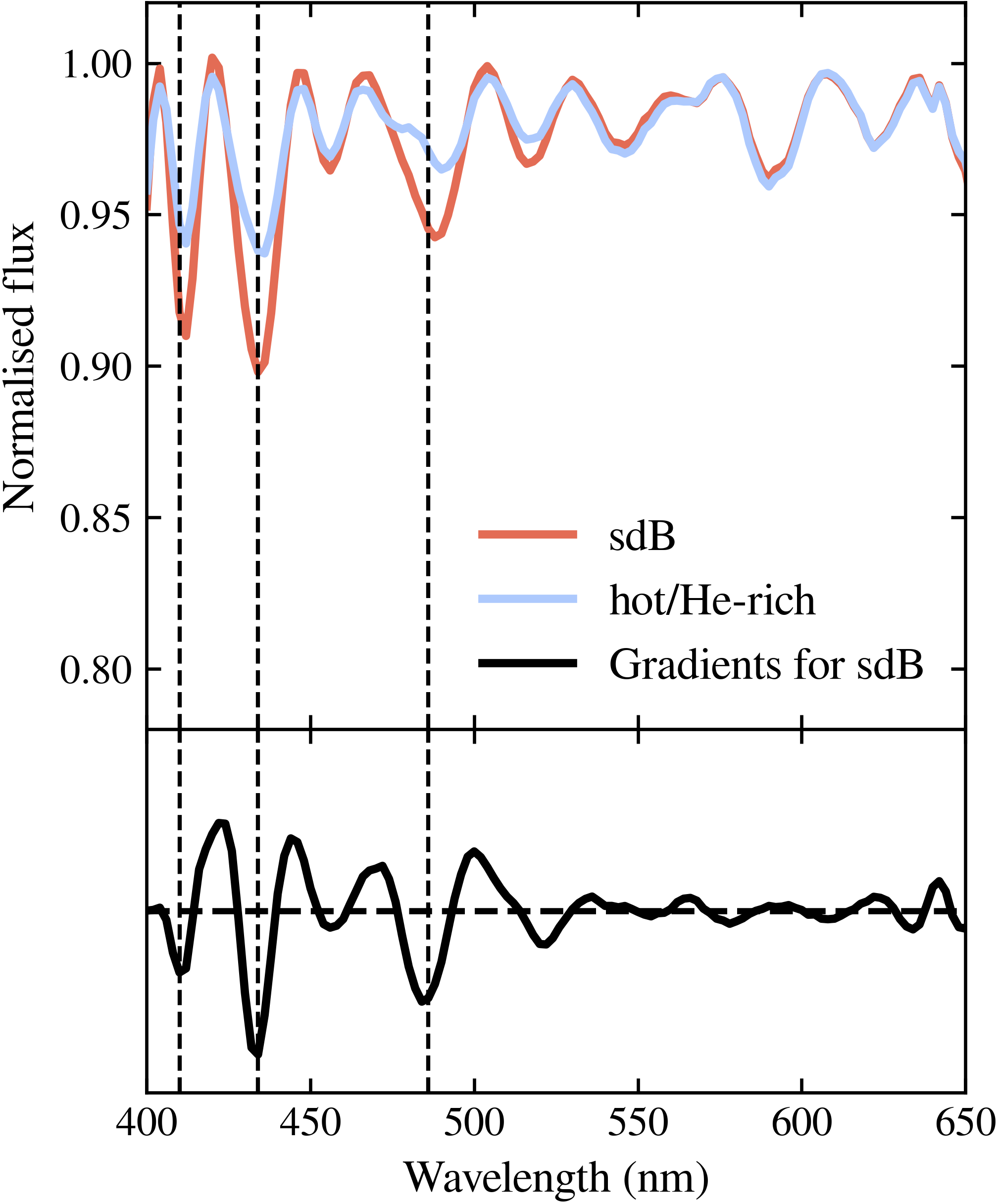}
  \caption{Top panel: Mean spectra of the classes sdB (red) and hot/He-rich (blue) in the CNN training set. Bottom panel: Network gradients for the class sdB. In both panels, dashed vertical lines show the positions of the 410, 434, 486~nm Balmer absorption lines of hydrogen.}
  \label{figure: CNN_gradients_sdB}
\end{figure}

\begin{figure}[!t] \centering \includegraphics[width=1.1\columnwidth]{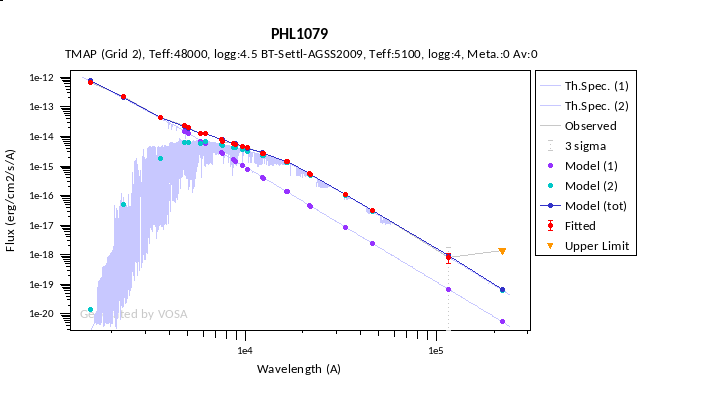} \includegraphics[width=1.1\columnwidth]{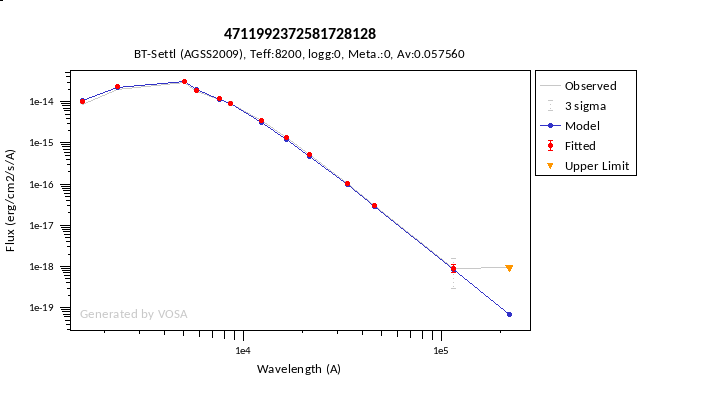} \caption{Examples of VOSA SED fits. Top panel: Fit of the binary hot hds PHL1079 (Gaia DR3 2560685069816099968). For a good fit, the observed SED (red data points) requires a combination of a TMAP model of a hot star (Model 1, purple data points) and a BT-Settl model of a cooler main sequence star (Model 2, cyan data points). Theoretical spectra for both models are shown in lavender colour. The combined model is shown in blue. Bottom panel: Fit of a main sequence BT-Settl model SED (blue) to the observed SED (red) of the island A object with Gaia DR3 4711992372581728128. In both panels, the parameters of the fitted models are given at the top of the plot. The yellow triangles mark upper limits, and are excluded from the model fitting.} \label{figure: VOSA_fits} \end{figure}

\begin{figure*}[!t]
  \centering
  \includegraphics[width=\textwidth]{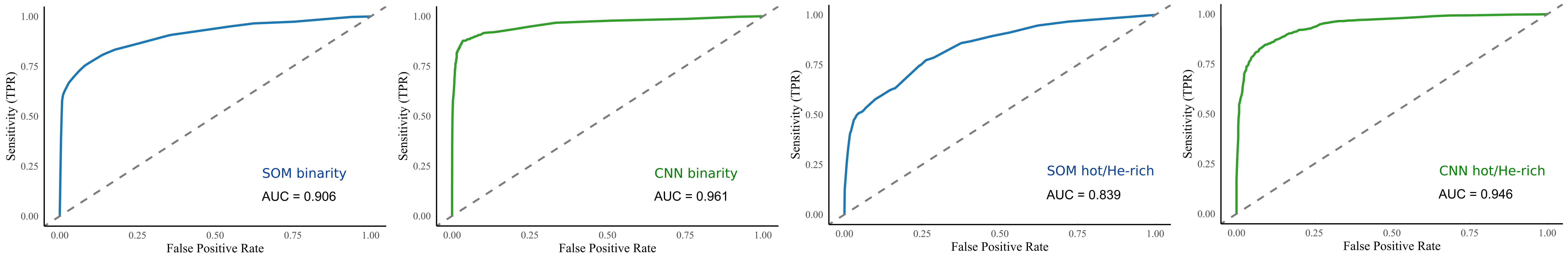}
  \caption{ROC curves for different models and classification tasks.}
  \label{figure: ROC_curves}
\end{figure*}

\begin{figure*}[!b]
  \centering
  \includegraphics[width=\textwidth]{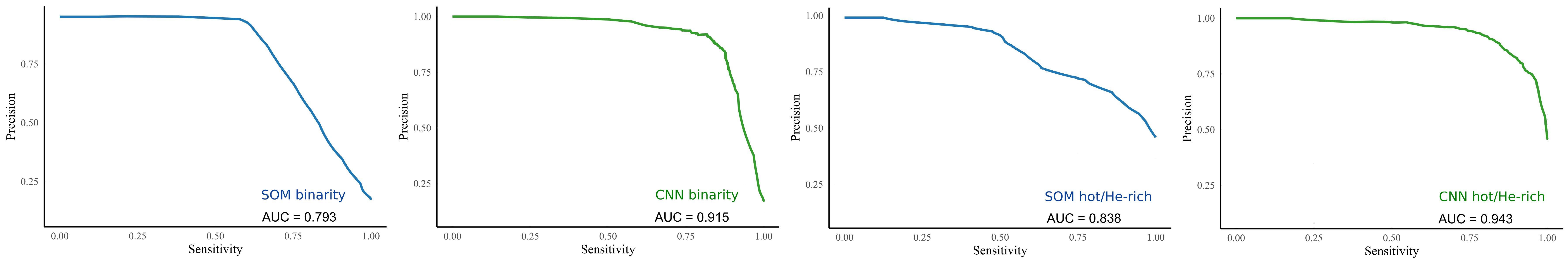}
  \caption{PR curves for different models and classification tasks.}
  \label{figure: PR_curves}
\end{figure*}

\begin{figure*}[!t]
  \centering
  \includegraphics[width=0.8\textwidth]{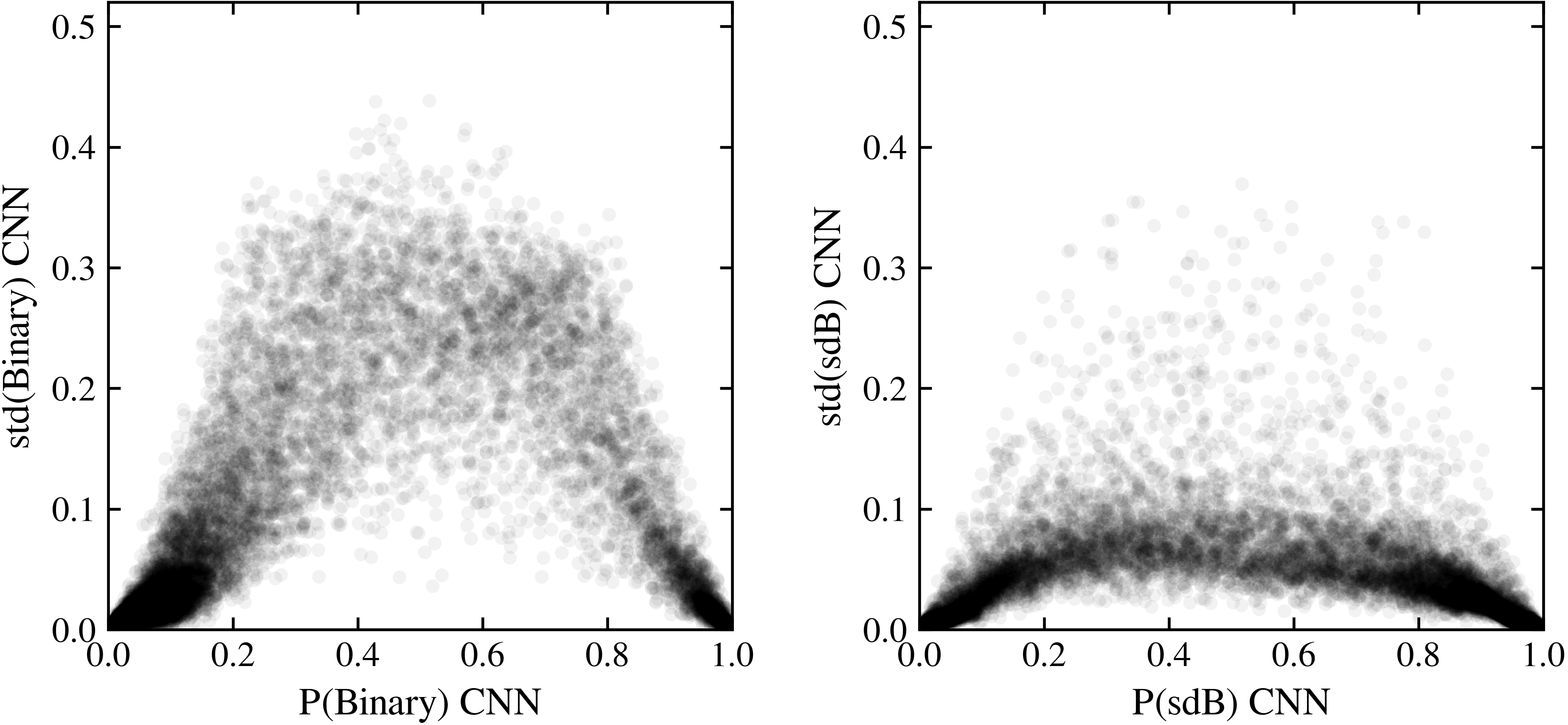}
  \caption{Left panel: Internal CNN uncertainties relative to predicted binary probability for our sample of 20061 hot sds. Right panel: Internal CNN uncertainties relative to predicted sdB probability. Darker shade means higher density of data points.}
  \label{figure: CNN_uncertainties}
\end{figure*}

\FloatBarrier
\begin{table*}
\caption{Classification probabilities for binary and hot/He-rich subdwarfs.}
\centering
\begin{tabular}{cccccccc}
\hline
\hline
\noalign{\smallskip}
GaiaEDR3 & \makecell{P(binary)\\CNN} & \makecell{std(binary)\\CNN} & \makecell{P(sdB)\\CNN} & \makecell{std(sdB)\\CNN} & \makecell{P(binary)\\SOM} & \makecell{P(hot/He-rich)\\SOM} & UMAP region \\
\hline
\noalign{\smallskip}
1306361548360576 & 0.036 & 0.012 & 0.956 & 0.010 & 0.078 & 0.286 & Main body \\
\noalign{\smallskip}
1792620565667968 & 0.021 & 0.009 & 0.914 & 0.019 & 0.078 & 0.152 & Main body \\
\noalign{\smallskip}
6052403489630720 & 0.044 & 0.012 & 0.965 & 0.008 & 0.091 & 0.211 & Main body \\
\noalign{\smallskip}
6353119919810816 & 0.136 & 0.076 & 0.076 & 0.022 & 0.030 & 0.079 & Main body \\
\noalign{\smallskip}
10844075163628928 & 0.069 & 0.021 & 0.099 & 0.027 & 0.045 & 0.036 & Main body \\
\noalign{\smallskip}
11015044926034816 & 0.042 & 0.013 & 0.306 & 0.057 & 0.000 & 0.063 & Main body \\
\noalign{\smallskip}
11963171843658240 & 0.074 & 0.011 & 0.480 & 0.046 & 0.250 & 0.211 & Main body \\
\noalign{\smallskip}
... & ... & ... & ... & ... & ... & ... & ... \\
\hline
\end{tabular}
\label{table: results_CNN_SOM}
\end{table*}

\end{appendix}

\end{document}